\documentclass[sor,jor,reprint,amsmath,amssymb,longbibliography,article-title]{revtex4-2}

\usepackage{graphicx}
\usepackage{bm}
\usepackage{booktabs}
\usepackage{siunitx}
\usepackage{microtype}
\usepackage{placeins}
\usepackage{hyperref}
\hypersetup{hidelinks}

\newcommand{\dd}{\mathrm{d}}
\newcommand{\tr}{\operatorname{tr}}
\newcommand{\dev}{\operatorname{dev}}
\newcommand{\LR}{\mathrm{LR}}
\newcommand{\vix}{\mathrm{vix}}
\newcommand{\CL}{\mathrm{CL}}
\newcommand{\CCR}{\mathrm{CCR}}
\newcommand{\eff}{\mathrm{eff}}
\newcommand{\rec}{\mathrm{rec}}
\newcommand{\rel}{\mathrm{rel}}

\begin{document}

\title{Generalized Constraint-Controlled Rheological Model: Two Constitutive Pathways under Sinusoidal Periodic Shear Deformation}

\author{Dario Nichetti}
\email{dario.nichetti@rheonicsrl.com}
\affiliation{Rheonic Lab, Via Quadelle 2C, 26012 Castelleone (CR), Italy}

\author{Alessio Zaccone}
\email{alessio.zaccone@unimi.it}
\affiliation{Department of Physics ``Aldo Pontremoli'', University of Milan, Via Celoria 16, 20133 Milan, Italy}

\date{\today}

\begin{abstract}
We formulate the Generalized Constraint-Controlled Rheological Model (GCCRM) for homogeneous simple shear under sinusoidal periodic shear deformation (SPSD), from the linear viscoelastic regime to strongly nonlinear response in entangled linear and long-arm-branched polymer melts. Molecular architecture, conformation, chain stretch, flow-active load-bearing constraints, convective constraint release, the associated flow-induced dynamic-tube state, and activated arm retraction define one common upstream molecular state. That state feeds two time-domain constitutive pathways: a direct configurational projection and a driven dissipative internal-variable pathway, which are recombined into the total shear stress before Fourier analysis. The long-arm-branched architecture is specified by an entire backbone of molecular weight $M_{bb}$ and $N_a$ long arms of molecular weight $M_a$; the branch partition is a compliance-weighted representative-constitutive-cell (RCC) closure, not a local T-junction identity. The configurational stress is normalized so that the two-population linear modulus sums to the entanglement modulus without a second architecture weighting. Dynamic tube dilation and stretch enter the dissipative activation through distinct molecular drivers. Under the stated reduced closures, GCCRM yields a smooth elastic response-class map containing a molecular onset scale, an approximately logarithmic growth window, a branch-dependent maximum, and distinct high-deformation asymptotes. Fourier harmonics, elastic nonlinearity ratio (ELR), viscous nonlinearity ratio (VLR), and Lissajous representations are post-processing observables of the total SPSD stress. The present work establishes the constitutive architecture and its analytical consistency; quantitative predictive validation requires a single parameter set per material tested across SPSD and, prospectively, transient shear histories.
\end{abstract}

\maketitle


\section{Introduction}

The response of an entangled polymer melt under sinusoidal periodic shear deformation (SPSD) reflects a competition between molecular orientation and stretch, relaxation of the confining tube, convective constraint release (CCR), dynamic tube dilation, and, in long-chain-branched architectures, activated arm retraction. Dynamic dilution of the tube in branched and polydisperse entangled systems has been developed theoretically and tested experimentally across a broad literature.~\cite{BallMcLeish1989,Watanabe2008,vanRuymbeke2006,Das2006,vanRuymbeke2014,Gold2019,MaloDeMolina2019} The forcing considered throughout this work is
\begin{equation}
\gamma(t)=\gamma_0\sin(\omega t),
\qquad
\dot\gamma(t)=\omega\gamma_0\cos(\omega t),
\label{eq:SPSDforcing}
\end{equation}
for arbitrary strain amplitude $\gamma_0$. Thus SPSD denotes the complete forcing protocol, including the linear viscoelastic regime, the linear-to-nonlinear crossover, and the strongly nonlinear regime; it is not restricted to large amplitudes.

The tube/primitive-chain framework of de Gennes and Doi--Edwards provides the molecular basis for entangled-polymer dynamics, while later constitutive theories incorporate contour-length fluctuations, stretch, and CCR under strong deformation.~\cite{deGennes1971,DoiEdwardsPart1,DoiEdwardsPart2,DoiEdwardsPart3,DoiEdwards1986,LikhtmanMcLeish2002,Graham2003,LikhtmanGraham2003,MeadLarsonDoi1998,McLeish2002} Long-chain branching introduces an additional architecture-dependent time scale because an entangled side arm can act first as a temporary anchor and subsequently relax by activated retraction.~\cite{MilnerMcLeish1997,MilnerMcLeishArm1998,McLeish2002,Watanabe1999,McLeishLarson1998,Blackwell2000,vanRuymbeke2014} A recent reduced response theory identified delayed arm retraction as a mechanism capable of generating the characteristic elastic maximum of long-chain-branched melts.~\cite{NichettiZaccone2026Delayed} A complementary constraint-limited tube-orientation theory describes the finite orientational response of entangled polymers under oscillatory shear deformation.~\cite{NichettiZaccone2026Constraint}

The purpose of the present work is to organize the Generalized Constraint-Controlled Rheological Model (GCCRM) around one molecular state rather than around separate ``elastic'' and ``viscous'' molecular models. Architecture, conformation, stretch, flow-active constraints, their associated dynamic-tube state, CCR, and arm retraction first generate a common upstream molecular trajectory. That trajectory feeds two constitutive pathways: a direct configurational projection $\tau_E(t)$ and a driven dissipative internal-variable pathway $\mathcal R_i(t)\to X_i(t)\to\tau_V(t)$. They are then recombined,
\begin{equation}
\tau(t)=\tau_E(t)+\tau_V(t),
\end{equation}
and only the total periodic stress is subsequently projected onto Fourier harmonics. This ordering is essential: the constitutive bifurcation and the Fourier sine/cosine decomposition are different operations. A configurational stress with memory can project onto both Fourier quadratures, and an intra-cycle-modulated dissipative stress can do the same.

Previous Fourier-resolved work established the generalized elastic and viscous post-processing of a periodic stress waveform and the use of higher harmonics as sensitive measures of nonlinear viscoelasticity.~\cite{Wilhelm1998,Wilhelm2002,Cho2005,Ewoldt2008,Ewoldt2013,EwoldtBharadwaj2013,ScacchiNichetti2025RCT,ScacchiNichetti2025AJOP} In the present notation, the elastic nonlinearity ratio (ELR) and viscous nonlinearity ratio (VLR) quantify the weighted higher-odd-harmonic contributions relative to the corresponding first-harmonic storage and loss moduli, respectively. Their explicit definitions are given in Sec.~VI. Those derivations are not repeated here. Likewise, a Lissajous trajectory reconstructed from the same harmonics is a representation of the same SPSD signal rather than an independent constitutive observable. The new content is the molecular state dynamics that generate the total stress before this post-processing step.
The manuscript distinguishes six logical levels: definitions/identities, governing evolution equations, constitutive closures, asymptotic or reduced closures, derived consequences, and experimental observables. This terminology is used deliberately. A statement is called a prediction only when it follows from solving the model with parameters constrained independently of the datum being predicted. In the absence of a complete harmonic-level parameter-constrained validation, the present paper is therefore positioned as a constitutive framework whose governing equations and stated closures are capable of generating the response classes discussed below.

\section{Common molecular architecture and dynamics}
\subsection{Architecture and branch-point partition}

The molecular architecture is defined without an additional ``active-arm count''. The modeled long-arm-branched chain has an entire backbone of molecular weight $M_{bb}$ and $N_a$ long arms, each of molecular weight $M_a$,
\begin{equation}
M=M_{bb}+N_aM_a,
\label{eq:Mtotal}
\end{equation}
where $N_a$ is the number of long arms in the architecture. The corresponding entanglement numbers are
\begin{equation}
Z_{bb}=\frac{M_{bb}}{M_e},
\qquad
Z_a=\frac{M_a}{M_e},
\label{eq:Zdefs}
\end{equation}
with $M_e$ the entanglement molecular weight. Physical mass/population fractions are
\begin{equation}
w_{bb}=\frac{M_{bb}}{M},
\qquad
w_a=\frac{N_aM_a}{M},
\qquad
w_{bb}+w_a=1,
\label{eq:weights}
\end{equation}
and the architecture loading ratio is
\begin{equation}
\Xi\equiv\frac{N_aZ_a}{Z_{bb}}
=\frac{N_aM_a}{M_{bb}}.
\label{eq:Xi}
\end{equation}
Thus $\Xi$ measures the total long-arm loading relative to the entire backbone, while the absolute arm entanglement number $Z_a$ remains a distinct dynamical quantity because it controls activated arm retraction.

\begin{figure*}[t]
\centering
\includegraphics[width=0.86\textwidth]{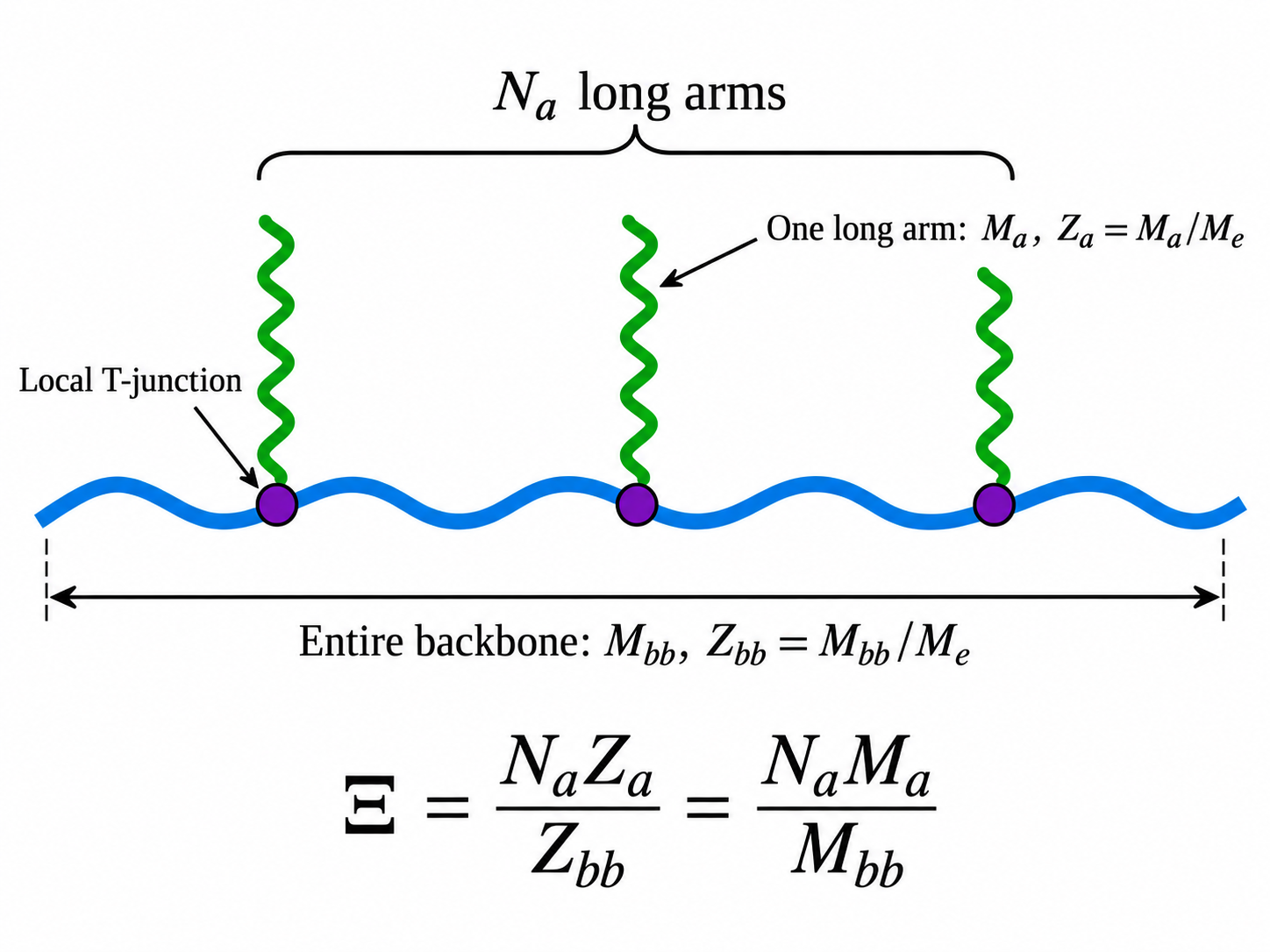}
\caption{Molecular architecture used in GCCRM. The blue chain represents the entire backbone, of molecular weight $M_{bb}$ and entanglement number $Z_{bb}=M_{bb}/M_e$. The molecule carries $N_a$ long arms, each of molecular weight $M_a$ and entanglement number $Z_a=M_a/M_e$. Each physical branch point remains a local three-leg T-junction. The architecture ratio $\Xi=N_aZ_a/Z_{bb}=N_aM_a/M_{bb}$ measures the total arm loading relative to the entire backbone, whereas the absolute arm entanglement number $Z_a$ separately controls the activated arm-retraction time.}
\label{fig:architecture}
\end{figure*}

For one physical T-junction, local three-leg force balance and displacement compatibility establish mechanical coupling but do not determine a quantitative backbone--arm partition. GCCRM therefore introduces at the representative-constitutive-cell (RCC) level the explicit coarse-grained compliance closure
\begin{equation}
C_{bb}=C_eZ_{bb},
\qquad
C_a^{\mathrm{hom}}=N_aC_eZ_a,
\label{eq:RCCcomplianceMain}
\end{equation}
which, together with compliance-weighted redistribution of the stored generalized configurational displacement, gives
\begin{equation}
\Phi_{bb}^{0}=\frac{Z_{bb}}{Z_{bb}+N_aZ_a}=\frac{1}{1+\Xi},
\qquad
\Phi_a^{0}=\frac{N_aZ_a}{Z_{bb}+N_aZ_a}=1-\Phi_{bb}^{0}.
\label{eq:Phi0}
\end{equation}
Equation~\eqref{eq:Phi0} is therefore a constitutive result conditional on local force balance, local compatibility, Gaussian coarse-grained compliance scaling, and RCC homogenization over the $N_a$ arms; it is not an identity derived from the local T-junction geometry. Appendix~\ref{app:partition} gives the derivation. Importantly, the equilibrium quantities $\Phi_i^0$ represent the configurational displacement/energy partition generated by this closure. The mass fractions $w_i$ remain useful physical population fractions, particularly in the dissipative pathway, but are not applied as a second configurational weighting. For the simplified equal-$M_e$ architecture of Eqs.~\eqref{eq:weights} and \eqref{eq:Phi0}, $w_i$ and $\Phi_i^0$ happen to have the same numerical values; they nevertheless have different constitutive meanings and are not multiplied together in the configurational stress.

\subsection{Common molecular state}

For each molecular population $i\in\{bb,a\}$, the configuration is represented by a symmetric positive-definite conformation tensor $\bm A_i$. The stretch ratio and traceless normalized orientation are
\begin{equation}
\lambda_i(t)=\left[\frac{\tr\bm A_i(t)}{3}\right]^{1/2},
\qquad
\bm S_i^{u}(t)=\frac{\bm A_i(t)}{\tr\bm A_i(t)}-\frac{\bm I}{3}.
\label{eq:lambdaS}
\end{equation}
The superscript $u$ denotes the conformation-predicted orientation before application of the architecture-dependent constraint-limited mapping. In the present closure, $\nu_i(t)$ is the \emph{flow-active load-bearing constraint fraction} associated specifically with the CCR/recovery kinetics of Eq.~\eqref{eq:nu}. It is not identified with the total microscopic entanglement population or with the complete physical dynamic dilution of the surrounding tube. Its physical interval is $0\le\nu_i\le1$, with quiescent normalization $\nu_i=1$. Other relaxation processes, including backbone disengagement, arm retraction, matrix relaxation, and stretch relaxation, act elsewhere in the model rather than as additional direct sinks of $\nu_i$.

The minimal common molecular state is therefore generated from
\begin{equation}
\mathcal Z_i^{(0)}(t)=\{\bm A_i,\lambda_i,\bm S_i^u,\nu_i\},
\label{eq:minstate}
\end{equation}
before derived tube-state quantities are introduced.

\subsection{Dynamic tube state and constraint-limited orientation}

Dynamic tube dilation is represented by
\begin{equation}
a_i(t)=a_{0,i}\nu_i(t)^{-\alpha_i/2},
\qquad
Z_i^{\eff}(t)=Z_i\nu_i(t)^{\alpha_i},
\label{eq:dilation}
\end{equation}
where $a_{0,i}$ is the equilibrium tube diameter and $\alpha_i\ge0$ is the GCCRM flow-induced dynamic-dilution exponent. Equations~\eqref{eq:dilation} are a closure for the tube state associated with the flow-active constraint population $\nu_i$; they are not asserted to represent every contribution to physical dynamic tube dilation, which in tube theories can also reflect relaxation of the surrounding matrix, arm relaxation, and constraint release.~\cite{BallMcLeish1989,Watanabe2008,vanRuymbeke2006,Das2006,Gold2019,MaloDeMolina2019} The instantaneous architecture partition becomes
\begin{equation}
\Phi_{bb}(t)=
\frac{Z_{bb}^{\eff}(t)}{Z_{bb}^{\eff}(t)+N_aZ_a^{\eff}(t)},
\qquad
\Phi_a(t)=1-\Phi_{bb}(t).
\label{eq:Phi}
\end{equation}
This establishes the causal chain $\nu_i\rightarrow Z_i^{\eff}\rightarrow\Phi_i$.

The conformation-predicted shear orientation is mapped to the bounded constraint-limited value
\begin{equation}
S_{xy,i}^{\CL}(t)
=\frac{1}{2}
\tanh\!\left[\Phi_i(t)\,\operatorname{artanh}\!\left(2S_{xy,i}^{u}(t)\right)\right].
\label{eq:CLmap}
\end{equation}
Positive definiteness of $\bm A_i$ already implies
\begin{equation}
|S_{xy,i}^{u}|=\left|\frac{A_{xy,i}}{\tr\bm A_i}\right|<\frac12.
\label{eq:geomBound}
\end{equation}
For a strictly positive-definite conformation tensor, $1/2$ is therefore an asymptotic limiting value rather than an attainable equality. The tanh--artanh mapping does not create this geometric bound; it preserves it while introducing architecture- and constraint-dependent attenuation through $\Phi_i$. At small orientation, $S_{xy,i}^{\CL}\simeq\Phi_iS_{xy,i}^{u}$; in the linear-chain backbone limit, $\Phi_{bb}=1$ gives $S_{xy,bb}^{\CL}=S_{xy,bb}^{u}$. The specific mapping in Eq.~\eqref{eq:CLmap} remains a GCCRM functional closure. Its energetic status is discussed separately in Sec.~\ref{sec:thermo} and Appendix~\ref{app:thermo}.

The equilibrium quantities $\Phi_i^0$ have a strict partition meaning within the quadratic RCC compliance closure: they partition the generalized configurational displacement and the corresponding quadratic stored configurational energy. At finite deformation, the dynamically updated $\Phi_i(t)$ entering the constraint-limited orientation mapping is used as a constraint-attenuation variable. Because the mapping is nonlinear, no exact finite-deformation identity of the form $\tau_{E,i}/\sum_j\tau_{E,j}=\Phi_i$ is assumed. The equilibrium partition is recovered to first order because $S_{xy,i}^{\CL}=\Phi_iS_{xy,i}^{u}+O[(S_{xy,i}^{u})^3]$.

\subsection{Relaxation hierarchy, CCR and constraint renewal}

The equilibrium backbone disengagement time is $\tau_{d,bb}^{0}$. Under dynamic dilution,
\begin{equation}
\tau_{d,bb}^{\eff}(t)
=\tau_{d,bb}^{0}
\left[\frac{Z_{bb}^{\eff}(t)}{Z_{bb}}\right]^{\mu_d}
=\tau_{d,bb}^{0}\nu_{bb}(t)^{\alpha_{bb}\mu_d}.
\label{eq:taud}
\end{equation}
For an entangled arm, the equilibrium activated retraction time is written
\begin{equation}
\tau_{a,\mathrm{ret}}^{0}
=\tau_e C_a Z_a^{\mu_a}\exp(\eta_aZ_a),
\label{eq:armtime0}
\end{equation}
where $\tau_e$ is the entanglement time, $C_a$ is a dimensionless arm-retraction prefactor, $\mu_a$ is the algebraic arm-length exponent, and $\eta_a$ is the activation coefficient controlling the exponential dependence on $Z_a$. This activated form is the primary arm-retraction time scale used by GCCRM and is consistent with established arm-retraction physics.~\cite{MilnerMcLeish1997,MilnerMcLeishArm1998,McLeish2002,Tezel2009} The dynamically diluted counterpart is
\begin{equation}
\tau_{a,\mathrm{ret}}^{\eff}(t)
=\tau_e C_a
\left[Z_a^{\eff}(t)\right]^{\mu_a}
\exp\!\left[\eta_aZ_a^{\eff}(t)\right].
\label{eq:armtime}
\end{equation}
We use the common notation
\begin{equation}
\tau_{\rel,bb}^{\eff}\equiv\tau_{d,bb}^{\eff},
\qquad
\tau_{\rel,a}^{\eff}\equiv\tau_{a,\mathrm{ret}}^{\eff},
\label{eq:taurel}
\end{equation}
throughout both constitutive pathways.

Under SPSD, the rate-of-deformation tensor is $\bm D=(\bm\kappa+\bm\kappa^T)/2$. The CCR driver is evaluated from the conformation-predicted orientation,
\begin{equation}
\Gamma_{\CCR,i}(t)
=\beta_i\left|\bm D(t):\bm S_i^u(t)\right|,
\qquad \beta_i\ge0,
\label{eq:CCR}
\end{equation}
which places CCR upstream of the constraint-limited stress projection and follows the molecular picture of flow-induced constraint release used in tube-based constitutive models.~\cite{MilnerMcLeishLikhtman2001,Marrucci1996,IannirubertoMarrucci2014,Baig2010,Yaoita2008,Furuichi2010} Using $\bm S_i^{\CL}$ instead would define a different constitutive closure and is not mixed with Eq.~\eqref{eq:CCR}.

The active-constraint population obeys
\begin{equation}
\frac{\dd\nu_i}{\dd t}
=-\Gamma_{\CCR,i}(t)\nu_i
+\frac{1-\nu_i}{\tau_{\rec,i}},
\qquad \tau_{\rec,i}>0.
\label{eq:nu}
\end{equation}
Equation~\eqref{eq:nu} is a governing kinetic equation with a specified CCR/recovery closure. It restores $\nu_i=1$ as the quiescent fixed point and avoids counting quiescent disengagement or arm retraction a second time as direct destruction of $\nu_i$. The physical interval is invariant: at $\nu_i=0$, $\dot\nu_i=\tau_{\rec,i}^{-1}>0$, whereas at $\nu_i=1$, $\dot\nu_i=-\Gamma_{\CCR,i}\le0$. Hence initial data in $[0,1]$ remain in $[0,1]$.
\subsection{Objective evolution of the common state}

The upper-convected derivative is
\begin{equation}
\overset{\nabla}{\bm A}_i
\equiv
\frac{\dd\bm A_i}{\dd t}
-\bm\kappa\cdot\bm A_i
-\bm A_i\cdot\bm\kappa^T.
\label{eq:ucd}
\end{equation}
The finite-stretch factor is taken as
\begin{equation}
F_i(\lambda_i)=
\frac{1-\lambda_i^{-1}}{1-\lambda_i/\lambda_{\max,i}},
\qquad 0<\lambda_i<\lambda_{\max,i},
\label{eq:finite}
\end{equation}
with $\lambda_{\max,i}>1$. The common conformation dynamics are
\begin{equation}
\overset{\nabla}{\bm A}_i
=-\frac{\bm A_i-\bm I}{\tau_{\rel,i}^{\eff}}
-2\Gamma_{\CCR,i}\,\dev\bm A_i
-\frac{2}{\tau_{R,i}^{\eff}}F_i(\lambda_i)\bm A_i.
\label{eq:Acommon}
\end{equation}
The flow-assisted stretch relaxation time is completed by the constitutive closure
\begin{equation}
\frac{1}{\tau_{R,i}^{\eff}(t)}
=\frac{1}{\tau_{R,i}^{0}}
+\beta_{s,i}\left|\bm D:\bm S_i^u\right|,
\qquad \beta_{s,i}\ge0.
\label{eq:tauR}
\end{equation}
The driver $|\bm D:\bm S_i^u|$ is chosen because it measures deformation-rate work on the conformation-predicted orientation already present in the common state and therefore vanishes in the absence of flow-induced orientational activity. A closure based directly on $|\dot\gamma|$ would not distinguish deformation that is ineffective at orienting the molecular state, while a closure based on $\bm S_i^{\CL}$ would move architecture-specific stress partitioning upstream into the common dynamics. Equation~\eqref{eq:tauR} is therefore a model closure, not a definition.

The scalar conformation magnitude $\lambda_i=(\tr\bm A_i/3)^{1/2}$ is not constrained to remain above unity during a shear-reversal cycle. Taking the trace of Eq.~\eqref{eq:Acommon} gives
\begin{equation}
6\lambda_i\dot\lambda_i
=2\bm D:\bm A_i
-\frac{3(\lambda_i^2-1)}{\tau_{\rel,i}^{\eff}}
-\frac{6\lambda_i^2}{\tau_{R,i}^{\eff}}F_i(\lambda_i).
\label{eq:lambdatrace}
\end{equation}
In simple shear, $\bm D:\bm A_i=\dot\gamma A_{xy,i}$, and at $\lambda_i=1$ one obtains $\dot\lambda_i=\dot\gamma A_{xy,i}/3$, which can be negative after flow reversal. The admissible stretch domain is therefore taken as $0<\lambda_i<\lambda_{\max,i}$. Values below unity represent transient contraction of the scalar conformation magnitude below its equilibrium value. The adopted stretch function changes sign below unity and drives recovery toward $\lambda_i=1$, while retaining the finite-extensibility barrier at $\lambda_{\max,i}$.

Equations~\eqref{eq:dilation}--\eqref{eq:Acommon} generate the common molecular trajectory. The complete state available at the constitutive bifurcation is
\begin{equation}
\mathcal Z_i(t)=\{\bm A_i,\lambda_i,\bm S_i^u,S_{xy,i}^{\CL},\nu_i,
Z_i^{\eff},\Phi_i,\tau_{\rel,i}^{\eff},\Gamma_{\CCR,i},\tau_{R,i}^{\eff}\}.
\label{eq:fullstate}
\end{equation}
No uniqueness claim is made here for the full nonlinear periodic solution of this coupled state system; the phrase ``the molecular trajectory'' refers to the trajectory generated by the specified governing equations and initial/boundary data.

\subsection{Kinematic scope}

The constitutive construction is formulated here for homogeneous simple shear. The SPSD results use periodic steady state, but the time-domain governing equations themselves can be integrated under a nonperiodic shear-rate history without resetting the internal state. This observation motivates the cross-protocol transient validation proposed in Sec.~\ref{sec:transient}; it does not constitute a completed transient-flow validation. A general arbitrary-three-dimensional extension would still require an objective tensorial constraint-limited mapping and a corresponding full tensorial stress construction.

\section{Constitutive projections of the common state}
\subsection{Configurational stress pathway}

The first constitutive pathway is a direct projection of the common upstream state onto a configurational shear stress. With $G_e$ identified with the entanglement/plateau modulus, the stress normalization is
\begin{equation}
\tau_E(t)=3G_e\sum_{i\in\{bb,a\}}
\nu_i\lambda_i^2S_{xy,i}^{\CL}.
\label{eq:elasticstress}
\end{equation}
The factor 3 follows from the normalization $\lambda_i^2S_{xy,i}^{u}=A_{xy,i}/3$ and restores the conventional conformation-tensor normalization in the linear limit.~\cite{DoiEdwardsPart2,DoiEdwardsPart3,Giesekus1982} No mass fraction $w_i$ appears in Eq.~\eqref{eq:elasticstress}: the equilibrium RCC partition is already represented by $\Phi_i^0$, while its dynamically updated counterpart $\Phi_i(t)$ enters the nonlinear mapping as a constraint-attenuation variable; multiplying again by $w_i$ would introduce a second architecture weighting. The label ``configurational'' is intentional. Equation~\eqref{eq:elasticstress} is a time-domain stress contribution with memory and is not identified a priori with the sine Fourier quadrature. Its Fourier projection can contain both sine and cosine components. A stronger identification with a derivative of stored free energy would require the energetic completion discussed in Sec.~\ref{sec:thermo}.

\subsection{Dissipative stress pathway}

The second pathway is driven by the same upstream state $\mathcal Z_i(t)$ but passes through the downstream internal variable $X_i$. The upstream state first forms the non-negative molecular activity measure
\begin{equation}
\mathcal R_i
=c_{\CCR}R_{\CCR,i}
+c_{\mathrm{st}}R_{\mathrm{st},i}
+c_{\mathrm{td}}R_{\mathrm{td},i},
\label{eq:Rdrive}
\end{equation}
where
\begin{align}
R_{\CCR,i}&=\beta_i\tau_{\rel,i}^{\eff}|\dot\gamma|\,|S_{xy,i}^{u}|,
\label{eq:Rccr}\\
R_{\mathrm{st},i}&=\frac{\tau_{\rel,i}^{\eff}}{\tau_{R,i}^{\eff}}(\lambda_i-1)^2,
\label{eq:Rst}\\
R_{\mathrm{td},i}&=\left(1-\frac{Z_i^{\eff}}{Z_i}\right)^2
=\left(1-\nu_i^{\alpha_i}\right)^2.
\label{eq:Rtd}
\end{align}
The tube-dilation driver is therefore tied to the same active-constraint variable that defines $Z_i^{\eff}$ and is distinct from the stretch driver. Because only the products $k_{X,i}c_j$ enter Eq.~\eqref{eq:Xdyn}, the channel weights are normalized to remove a scale degeneracy,
\begin{equation}
c_{\CCR},c_{\mathrm{st}},c_{\mathrm{td}}\ge0,
\qquad
c_{\CCR}+c_{\mathrm{st}}+c_{\mathrm{td}}=1.
\label{eq:cdomains}
\end{equation}
The dissipative activation obeys its own first-order kinetic equation,
\begin{equation}
\frac{\dd X_i}{\dd t}
=k_{X,i}\mathcal R_i(t)(1-X_i)
-\frac{X_i}{\tau_{X,i}^{\rec}},
\qquad k_{X,i}\ge0,\quad \tau_{X,i}^{\rec}>0.
\label{eq:Xdyn}
\end{equation}
Thus $X_i$ has genuine internal dynamics. What it does not introduce is an independent molecular driving variable: its forcing $\mathcal R_i(t)$ is entirely determined by the history of the common GCCRM state. Appendix~\ref{app:Xmemory} proves causality, fading memory, invariance of $0\le X_i\le1$, and uniqueness of the periodic $X_i(t)$ conditional on a prescribed periodic $\mathcal R_i(t)$. That conditional result is not extended to uniqueness of the full coupled nonlinear molecular trajectory.

The population viscosity is
\begin{equation}
\eta_i(t)=\eta_{i,0}\left[1+A_VX_i^q(t)\right],
\qquad A_V\ge0,\quad q>0.
\label{eq:viscosity}
\end{equation}
Linear rheology constrains only the residual fast viscosity
\begin{equation}
\eta_f\equiv\sum_iw_i\eta_{i,0}.
\label{eq:etafdef}
\end{equation}
When the population activations $X_{bb}$ and $X_a$ differ, the individual baseline viscosities are not separately identifiable from $\eta_f$. We therefore introduce the normalized residual-viscosity partition
\begin{equation}
\varphi_{\eta,i}\equiv\frac{w_i\eta_{i,0}}{\eta_f},
\qquad \sum_i\varphi_{\eta,i}=1,
\label{eq:phieta}
\end{equation}
so that the primary dissipative stress law becomes
\begin{equation}
\tau_V(t)=\eta_f\dot\gamma(t)
\left[1+A_V\sum_{i\in\{bb,a\}}\varphi_{\eta,i}X_i^q(t)\right].
\label{eq:viscstress}
\end{equation}
For the minimal single-chemistry implementation used here, the additional closure $\eta_{bb,0}=\eta_{a,0}$ is adopted, giving $\varphi_{\eta,i}=w_i$. This equality is a parsimonious constitutive closure for the unresolved fast modes and is not claimed as a microscopic identity. For $\eta_f>0$ and $\varphi_{\eta,i}\ge0$ this pathway has non-negative instantaneous power. Like $\tau_E$, however, $\tau_V(t)$ is a time-domain waveform and is not identified a priori with the cosine Fourier quadrature; phase-dependent modulation of the dissipative activation can generate both Fourier quadratures.

\subsection{Non-redundancy of the dissipative channel and meaning of $\eta_{i,0}$}

CCR, stretch relaxation, and tube dilation already alter the common molecular state and thereby relax the configurational stress. Their appearance in $\mathcal R_i$ is not intended as a second copy of the same configurational stress relaxation. In Eqs.~\eqref{eq:Rdrive}--\eqref{eq:viscstress} they act as dimensionless indicators of molecular rearrangement activity that modulates a distinct irreversible momentum-transfer channel. No additional configurational stress is generated directly by $R_{\CCR,i}$, $R_{\mathrm{st},i}$, or $R_{\mathrm{td},i}$.

To avoid double counting already at linear order, $\eta_f$ is defined as a residual/fast-mode viscosity associated with residual/fast dissipative modes not represented by the two modeled conformation modes $\bm A_{bb}$ and $\bm A_a$. It is not the zero-shear viscosity of those same two entangled modes. The combination $\eta_f$ must be constrained from the residual linear loss after the two-population configurational response has been fixed; the normalized coefficients $\varphi_{\eta,i}$ then specify how that residual channel is partitioned between populations in the nonlinear law. The nonlinear factor $A_VX_i^q$ modulates only this distinct dissipative channel. A full microscopic energetic proof of orthogonality between the two bookkeeping channels is not claimed; the present formulation instead makes the separation explicit and testable.

\section{Linear-response consistency}
\label{sec:linear}

The complete model is linearized about
\begin{equation}
\bm A_i=\bm I,\qquad \lambda_i=1,\qquad \nu_i=1,\qquad X_i=0.
\label{eq:lineareq}
\end{equation}
Write $\bm A_i=\bm I+\bm a_i$ with $|\bm a_i|\ll1$. In homogeneous simple shear, $\Gamma_{\CCR,i}=O(\gamma_0^2)$, $1-\nu_i=O(\gamma_0^2)$, and $\lambda_i-1=O(\gamma_0^2)$, so CCR, dynamic dilution, stretch nonlinearity, and dissipative activation do not contribute at first order. The shear component of Eq.~\eqref{eq:Acommon} reduces to
\begin{equation}
\dot a_{xy,i}+\frac{a_{xy,i}}{\tau_{\rel,i}^{0}}=\dot\gamma,
\label{eq:linearAxy}
\end{equation}
where $\tau_{\rel,i}^{0}$ is the quiescent relaxation time. Since $S_{xy,i}^{u}=a_{xy,i}/3+O(\gamma_0^2)$ and $S_{xy,i}^{\CL}=\Phi_i^0a_{xy,i}/3+O(\gamma_0^2)$, the corrected configurational stress gives the modal modulus
\begin{equation}
G_i^0\equiv G_e\Phi_i^0,
\label{eq:Gi0}
\end{equation}
so that $\sum_iG_i^0=G_e$ because $\Phi_{bb}^0+\Phi_a^0=1$.
The resulting two-population reduced linear response is Maxwell-like,
\begin{align}
G_{\mathrm{conf}}'(\omega)&=\sum_iG_i^0\frac{(\omega\tau_{\rel,i}^{0})^2}{1+(\omega\tau_{\rel,i}^{0})^2},
\label{eq:Glinearprime}\\
G_{\mathrm{conf}}''(\omega)&=\sum_iG_i^0\frac{\omega\tau_{\rel,i}^{0}}{1+(\omega\tau_{\rel,i}^{0})^2}.
\label{eq:Glineardoubleprime}
\end{align}
At the same order, $X_i=0$ and the dissipative pathway contributes only the residual/fast term
\begin{equation}
\tau_V^{(0)}(t)=\eta_f\dot\gamma(t),
\label{eq:etafast}
\end{equation}
so that
\begin{equation}
G'(\omega)=G_{\mathrm{conf}}'(\omega),
\qquad
G''(\omega)=G_{\mathrm{conf}}''(\omega)+\omega\eta_f.
\label{eq:linearfull}
\end{equation}
Equations~\eqref{eq:Glinearprime}--\eqref{eq:linearfull} provide the required linear-response consistency check. The residual term $\omega\eta_f$ is the low-frequency representation of unresolved fast relaxation modes, not a frequency-independent replacement for an arbitrary missing spectrum. If the omitted modes have moduli $G_p$ and relaxation times $\tau_p$, then $G_p''=G_p\omega\tau_p/[1+(\omega\tau_p)^2]\simeq\omega G_p\tau_p$ only for $\omega\tau_p\ll1$. Accordingly, $G_{\mathrm{fast}}''=\omega\eta_f$, with $\eta_f=\sum_{p\in\mathrm{fast}}G_p\tau_p$, is restricted to the frequency window in which $\omega\tau_{\mathrm{fast,max}}\ll1$. Outside this window the residual contribution must be represented by an explicit multimode spectrum rather than by a Newtonian viscosity. The linearization also specifies the identification order: the two population relaxation times, configurational moduli, and residual fast viscosity must be constrained by linear rheology before nonlinear SPSD observables are used to test the remaining closures. The present $i\in\{bb,a\}$ model should not be described as a general resolved relaxation spectrum; a true multimode extension would require an additional mode index.

\section{Reduced nonlinear regime closures}
\subsection{Theoretical onset scale and smooth elastic regime map}

The theoretical molecular onset scale $\gamma_0^{\LR}$ is introduced through the cooperative-onset closure
\begin{equation}
\gamma_{0,\mathrm{lin}}^{\LR}
=C_{\LR}\left[Z_{bb}F_{\mathrm{CLF}}(Z_{bb})\right]^{-1/2},
\label{eq:onsetlin}
\end{equation}
where $F_{\mathrm{CLF}}(Z)$ is a positive dimensionless contour-length-fluctuation correction factor representing the coherent entanglement participation of a finite chain. No explicit functional form for $F_{\mathrm{CLF}}$ is used quantitatively in the present framework; a quantitative application must specify it from an independently chosen linear/tube model or reference.~\cite{MilnerMcLeish1998,LikhtmanMcLeish2002} When $F_{\mathrm{CLF}}$ varies slowly, Eq.~\eqref{eq:onsetlin} gives $\gamma_{0,\mathrm{lin}}^{\LR}\propto(M_e/M)^{1/2}$ in the highly entangled linear limit. Applying the equilibrium branch partition to the same closure gives
\begin{equation}
\gamma_{0,\mathrm{br}}^{\LR}
=\gamma_{0,\mathrm{lin}}^{\LR}(1+\Xi)^{1/2}.
\label{eq:onsetbr}
\end{equation}
These relations are constitutive/reduced closures, not kinematic identities. Experimentally, $\gamma_0^{\LR}$ should be estimated independently from the first experimentally resolvable higher odd harmonic above the instrumental and waveform noise floor, rather than by imposing a threshold on ELR or VLR. A formal statistical detection criterion would require a dedicated repeatability/noise protocol and is not defined in this work.

Define the reduced amplitude $x=\gamma_0/\gamma_0^{\LR}$. The approximate linearity of ELR versus $\log_{10}x$ is an intermediate-window behavior, not a piecewise law over the entire domain. For the smooth schematic regime map we therefore use the continuous reduced linear-backbone closure
\begin{equation}
Q_{\mathrm{lin}}(x)=Q_{\infty}
\frac{(x/x_s)^m}{1+(x/x_s)^m},
\qquad Q_{\infty}=3,
\label{eq:Qlinsmooth}
\end{equation}
with $x_s>0$ and $m>0$. The corresponding observable-level relation
\begin{equation}
\mathrm{ELR}_{\mathrm{lin}}(x)=\log_{10}[1+Q_{\mathrm{lin}}(x)]
\label{eq:ELRlinsmooth}
\end{equation}
is smooth, can display an approximately logarithmic growth window, and approaches $\log_{10}(4)$ asymptotically. The numerical value $\log_{10}(4)$ comes from the high-strain harmonic closure $Q_{\infty}=3$; it is not a consequence of the tanh--artanh orientation mapping alone.

\begin{figure*}[t]
\centering
\includegraphics[width=0.72\textwidth]{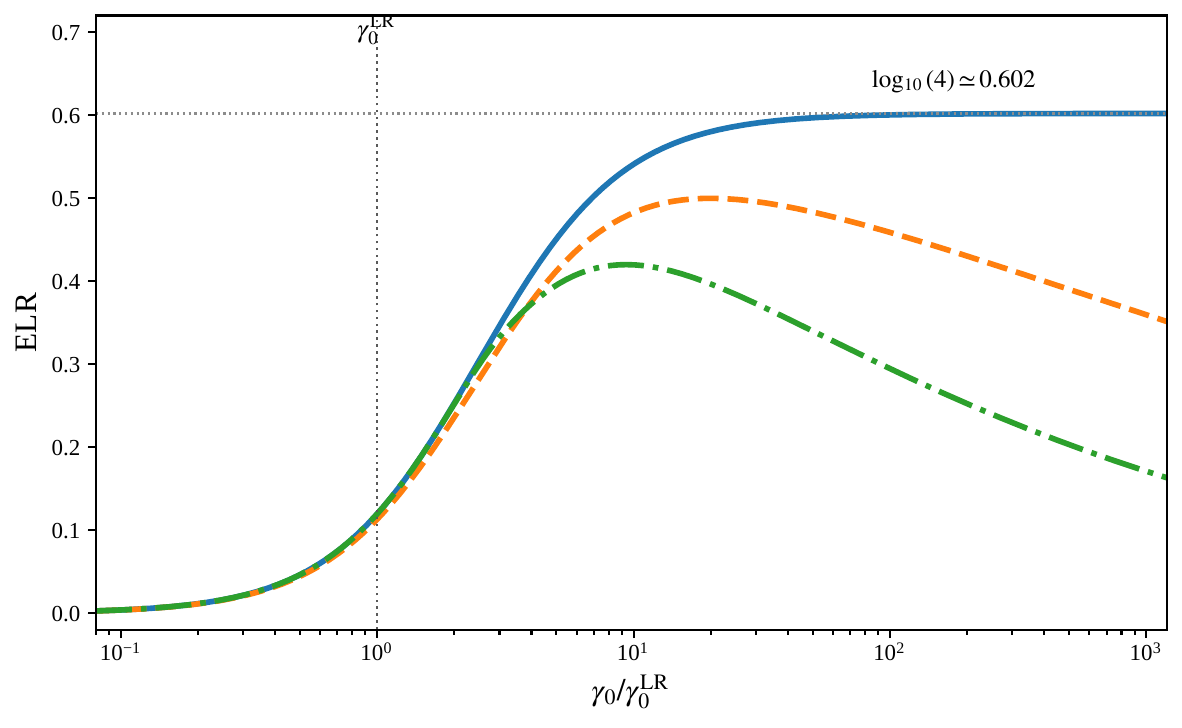}
\caption{Smooth schematic elastic response classes under SPSD after normalization by the theoretical onset scale $\gamma_0^{\LR}$. The continuous line is the idealized linear response class, the dashed line is a sparse-branched class, and the dash-dotted line is a more strongly branched class. All curves are generated from the continuous reduced closures in Eqs.~\eqref{eq:Qlinsmooth}--\eqref{eq:Qbrsmooth}; no piecewise-linear construction is used. The linear class approaches the separate high-strain closure $\mathrm{ELR}\to\log_{10}(4)$ smoothly. The branched maxima and post-maximum decay are generated by the reduced memory-survival closure $R_a$. In a quantitative application the crossover center is constrained by the primary arm-retraction dynamics through $\mathcal H_a(10^{u_r}\gamma_0^{\LR},\omega)\simeq1$ [Eq.~\eqref{eq:urhazard}], while $\kappa_a$ and $r_a$ remain reduced shape parameters. The figure is therefore a schematic phenomenological response-class map consistent with delayed arm retraction, not a fit and not a demonstration that the primary arm-retraction equations alone quantitatively generate the plotted maxima. The numerical value $\log_{10}(4)$ follows from the separate high-strain harmonic closure $Q_{\infty}=3$, not from the tanh--artanh orientation mapping.}
\label{fig:smoothregimes}
\end{figure*}

For the schematic branched response classes in Fig.~\ref{fig:smoothregimes}, we use a fully smooth delayed-memory closure in the logarithmic reduced coordinate
\begin{equation}
u\equiv\log_{10}x,
\qquad x=\frac{\gamma_0}{\gamma_0^{\LR}},
\label{eq:ulog}
\end{equation}
of the form
\begin{equation}
R_a(u;\Xi,u_r,\kappa_a,r_a)
=\left[1+\Xi\exp\!\left(\kappa_a(u-u_r)\right)\right]^{-r_a\ln(10)/\kappa_a}.
\label{eq:Rasmooth}
\end{equation}
Here $u_r$ is the center of the smooth memory-loss crossover, $\kappa_a>0$ controls its sharpness, and $r_a>0$ is the high-deformation memory-loss exponent. Equation~\eqref{eq:Rasmooth} is $C^{\infty}$: it approaches unity continuously for $u\ll u_r$ and behaves as $R_a\propto x^{-r_a}$ for $u\gg u_r$. The representative-cell factor $\Xi$ controls the strength of the architecture-dependent suppression. This relation is an explicit reduced GCCRM closure used to draw schematic response classes; it is not an independently established microscopic law and is not fitted to the experimental curves shown later.~\cite{NichettiZaccone2026Delayed}

For a quantitative material application, the crossover center $u_r$ should not remain disconnected from the primary activated arm-retraction dynamics. Because the dynamically diluted arm-retraction time varies within the SPSD cycle, a single instantaneous Deborah number does not uniquely define the reduced memory-loss crossover. We therefore introduce, only as a matching functional between the primary dynamics and the reduced response map, the cycle-integrated arm-retraction hazard
\begin{equation}
\mathcal H_a(\gamma_0,\omega)
=\int_0^T\frac{\dd t}{\tau_{a,\mathrm{ret}}^{\eff}(t;\gamma_0,\omega)}.
\label{eq:armhazard}
\end{equation}
For a first-order survival representation, the remaining arm-memory fraction after one cycle is
\begin{equation}
P_a(T)=\exp[-\mathcal H_a(\gamma_0,\omega)],
\label{eq:armhazardsurvival}
\end{equation}
and the reduced crossover center is matched to the primary arm-retraction dynamics through
\begin{equation}
\mathcal H_a(10^{u_r}\gamma_0^{\LR},\omega)\simeq1.
\label{eq:urhazard}
\end{equation}
This matching defines the location of the crossover but does not determine the reduced shape parameters $\kappa_a$ and $r_a$. The shorthand $\omega\tau_{a,\mathrm{ret}}^{\eff}\simeq1$ is recovered only when the intra-cycle variation of $\tau_{a,\mathrm{ret}}^{\eff}$ is weak. Consequently Fig.~\ref{fig:smoothregimes} is a schematic response-class map consistent with the arm-retraction mechanism, not a quantitative demonstration that the primary time-domain equations alone reproduce the plotted maximum.

The branched reduced distortion is then
\begin{equation}
Q_{\mathrm{br}}(x)=Q_{\mathrm{lin}}(x)R_a(x),
\qquad
\mathrm{ELR}_{\mathrm{br}}=\log_{10}[1+Q_{\mathrm{br}}(x)].
\label{eq:Qbrsmooth}
\end{equation}
The maximum is a derived consequence of the specified smooth closures, not a parameter-free theorem: because ELR is monotone in $Q_{\mathrm{br}}$, its position satisfies
\begin{equation}
\frac{\dd\ln Q_{\mathrm{lin}}}{\dd\ln x}
=-\frac{\dd\ln R_a}{\dd\ln x}.
\label{eq:peakcond}
\end{equation}
At large $x$, Eq.~\eqref{eq:Rasmooth} gives $R_a\propto x^{-r_a}$ and hence $Q_{\mathrm{br}}\to0$, while Eq.~\eqref{eq:Qlinsmooth} approaches $Q_{\infty}=3$. Figure~\ref{fig:smoothregimes} displays three smooth elastic response classes generated from Eqs.~\eqref{eq:Qlinsmooth}--\eqref{eq:Qbrsmooth}. The numerical values used solely to draw this schematic are reported explicitly in Appendix~\ref{app:hierarchy}, Table~\ref{tab:fig2params}.

Figure~\ref{fig:smoothregimes} contains only the elastic reduced response-class map. The separation $\gamma_{\vix}>\gamma_0^{\LR}$ is not represented by an artificial schematic curve; it is described in the text and shown directly by the experimental VLR data in Fig.~\ref{fig:experimental}(b).

\subsection{Dissipative activation, $\gamma_{\vix}$, and conditional VLR scaling}

The population-resolved Eq.~\eqref{eq:viscstress} is the primary dissipative law. The approximation
\begin{equation}
X_{bb}(t)\simeq X_a(t)\equiv X_V(t)
\label{eq:commonX}
\end{equation}
is only an optional reduced closure and cannot be used as independent evidence that normalized VLR must collapse across architectures.

When a scalar amplitude-domain activation is useful, it must be distinguished from the intra-cycle variable. We define
\begin{equation}
\overline X_{i,q}(\gamma_0,\omega)
\equiv
\left[\frac{1}{T}\int_0^T X_i^q(t;\gamma_0,\omega)\,\dd t\right]^{1/q}.
\label{eq:Xbar}
\end{equation}
The first molecular nonlinear scale is $\gamma_0^{\LR}$. A later experimentally observable crossover is denoted by
\begin{equation}
\gamma_{\vix}>\gamma_0^{\LR},
\label{eq:gammavix}
\end{equation}
where $\gamma_{\vix}$ is defined operationally as the onset of macroscopically resolved VLR growth relative to the experimental harmonic-resolution criterion. It is not a second microscopic transition. In the interval $\gamma_0^{\LR}<\gamma_0<\gamma_{\vix}$, nonlinear dissipative activity and higher harmonics may already be present while VLR remains close to the experimental resolution floor.

A reduced amplitude scaling such as
\begin{equation}
\overline X_{V,q}\propto
\left(\frac{\gamma_0}{\gamma_{\vix}}\right)^r
\label{eq:Xpower}
\end{equation}
does not by itself determine VLR. A logarithmic VLR branch follows only conditionally if the intra-cycle waveform admits an additional separable-shape closure, as shown in Appendix~\ref{app:fourierbridge}. The bound $0\le X_i\le1$ implies only
\begin{equation}
1\le 1+A_V\sum_i\varphi_{\eta,i}X_i^q(t)\le1+A_V,
\label{eq:etamax}
\end{equation}
not a universal nonzero VLR ceiling. If $X_i(t)\to1$ uniformly through the SPSD cycle, the viscosity becomes constant and the higher harmonics of the dissipative pathway vanish.
\section{Recombination and Fourier observables}
\subsection{Total stress}

The total shear stress is defined by the explicit constitutive decomposition
\begin{equation}
\tau(t)=\tau_E(t)+\tau_V(t).
\label{eq:totalstress}
\end{equation}
This is a testable constitutive choice, not a kinematic identity. The hierarchy is more precisely
\begin{equation}
\mathcal Z^{\mathrm{up}}(t)
\longrightarrow
\begin{cases}
\tau_E(t),\\
\mathcal R_i(t)\rightarrow X_i(t)\rightarrow\tau_V(t),
\end{cases}
\longrightarrow\tau(t)
\longrightarrow\{G_n',G_n''\}_{\mathrm{total}}.
\label{eq:hierarchychain}
\end{equation}
Thus one common upstream molecular state feeds a direct configurational projection and a driven dissipative internal-variable pathway. The final sine/cosine decomposition is spectral and must not be identified with the constitutive bifurcation.

\subsection{Fourier measurement bridge}

Only after periodic steady state under SPSD is reached is the total stress projected onto Fourier modes. The absence of even harmonics follows from the half-wave symmetry of the GCCRM solution. With $T=2\pi/\omega$, the imposed simple shear satisfies
\begin{equation}
\gamma(t+T/2)=-\gamma(t),
\qquad
\dot\gamma(t+T/2)=-\dot\gamma(t).
\label{eq:halfwaveforcing}
\end{equation}
For homogeneous simple shear, introduce the orthogonal transformation $\bm Q=\mathrm{diag}(-1,1,1)$. The GCCRM governing equations are equivariant under the simultaneous transformation $\bm\kappa\rightarrow\bm Q\bm\kappa\bm Q^T$ and $\bm A_i\rightarrow\bm Q\bm A_i\bm Q^T$. Consequently, in periodic steady state the shear components $S_{xy,i}^{u}$ and $S_{xy,i}^{\CL}$ change sign after one half-period, whereas the scalar variables $\lambda_i$, $\nu_i$, $Z_i^{\eff}$, $\Phi_i$, $\mathcal R_i$, and $X_i$ remain invariant. Hence both constitutive pathways obey
\begin{equation}
\tau_E(t+T/2)=-\tau_E(t),
\qquad
\tau_V(t+T/2)=-\tau_V(t),
\label{eq:halfwavestress}
\end{equation}
and therefore $\tau(t+T/2)=-\tau(t)$. All even-order Fourier coefficients consequently vanish, in agreement with the previously published SPSD Fourier treatment.~\cite{ScacchiNichetti2025AJOP}

With $\theta=\omega t$, the total periodic stress is therefore written as
\begin{equation}
\tau(\theta)=\gamma_0\sum_{n\,\mathrm{odd}}
\left[G_n'\sin(n\theta)+G_n''\cos(n\theta)\right].
\label{eq:FourierStress}
\end{equation}
The published generalized elastic and viscous procedures then define ELR and VLR from the total measured or calculated harmonic set.~\cite{Wilhelm1998,Wilhelm2002,Ewoldt2013,EwoldtBharadwaj2013,ScacchiNichetti2025RCT,ScacchiNichetti2025AJOP} For the ninth-order truncation used experimentally, the operational nonlinear combinations are
\begin{equation}
G_{\mathrm{NL}}'=3G_3'+5G_5'+7G_7'+9G_9',
\qquad
G_{\mathrm{NL}}''=-3G_3''+5G_5''-7G_7''+9G_9''.
\label{eq:GNLdefs}
\end{equation}
These compact definitions are quoted only to identify the observables; the published generalized-modulus derivation is not repeated. With the retained sign convention,
\begin{equation}
\mathrm{ELR}=\log_{10}\left(1-\frac{G_{\mathrm{NL}}'}{G_1'}\right),
\qquad
\mathrm{VLR}=\log_{10}\left(1+\frac{G_{\mathrm{NL}}''}{G_1''}\right).
\label{eq:ELRVLR}
\end{equation}
The detailed harmonic-weight/Chebyshev derivations are not repeated. Separate Fourier projections of $\tau_E$ and $\tau_V$ can be useful diagnostics, and Appendix~\ref{app:fourierbridge} gives the minimal model-specific bridge needed for the dissipative waveform, but the experimentally defined $G_n'$ and $G_n''$ in Eq.~\eqref{eq:ELRVLR} are projections of the total SPSD stress.

\section{Experimental connection and validation targets}

\begin{figure*}[!t]
\centering
\begin{minipage}[t]{0.49\textwidth}
\centering
\includegraphics[width=\linewidth]{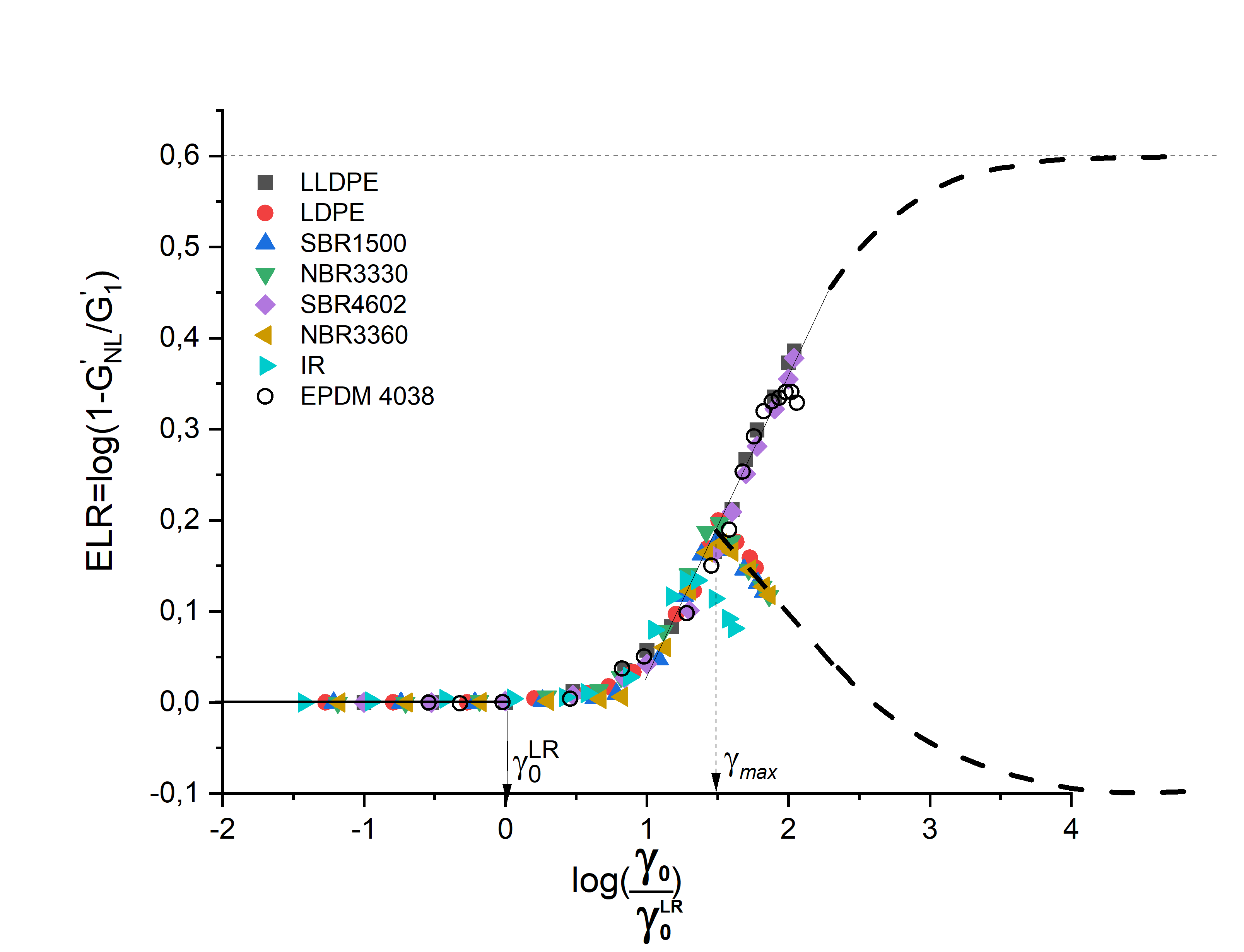}\\[-2pt]
\textbf{(a)}
\end{minipage}\hfill
\begin{minipage}[t]{0.49\textwidth}
\centering
\includegraphics[width=\linewidth]{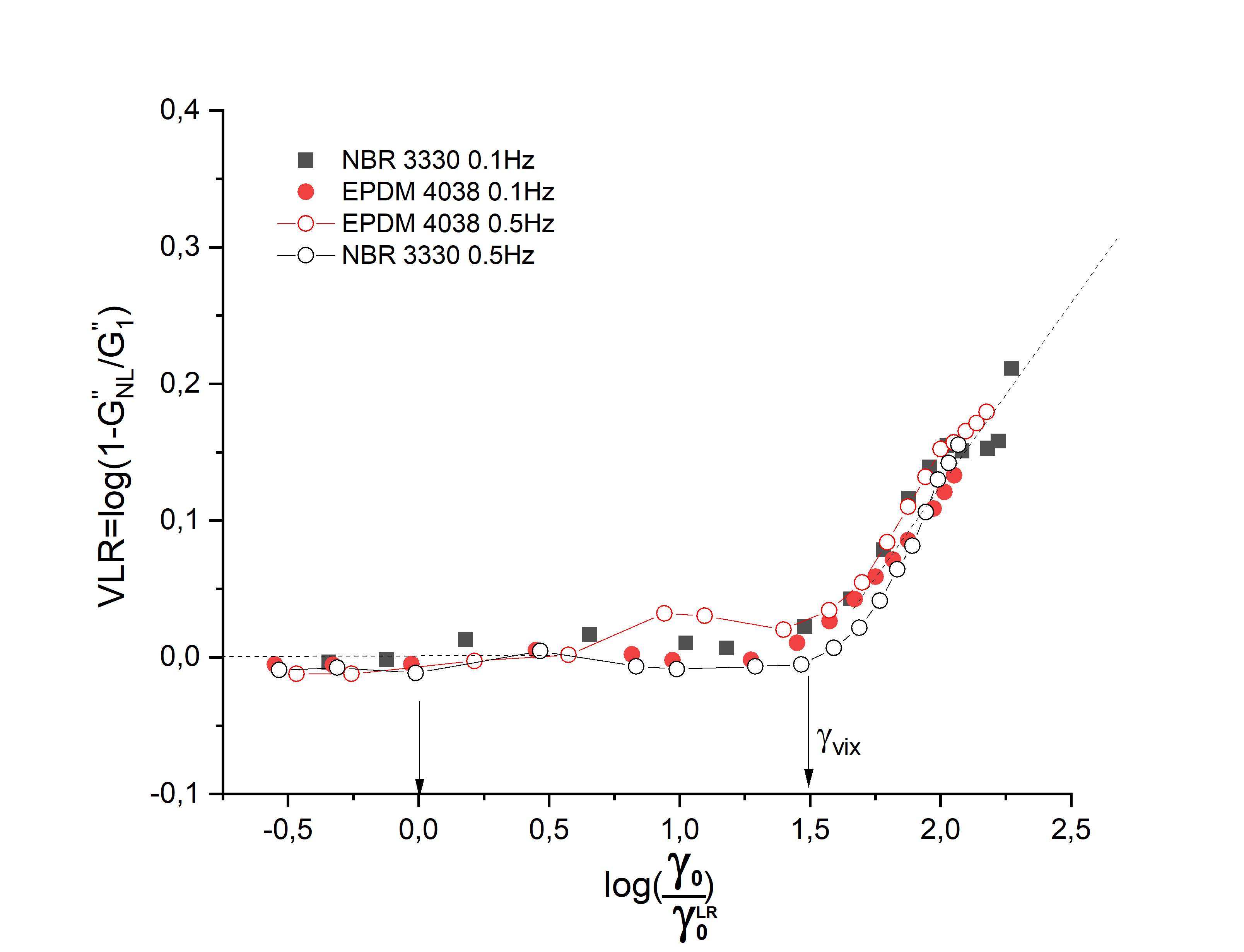}\\[-2pt]
\textbf{(b)}
\end{minipage}
\caption{Experimental SPSD observables used as validation targets.
\textbf{(a)} Elastic nonlinearity ratio (ELR) as a function of reduced
strain amplitude. The polymer datasets underlying the previously
reported nonlinear elastic response are replotted here in the ELR
representation from experimental data reported in
Refs.~\cite{ScacchiNichetti2025RCT,ScacchiNichetti2025AJOP} and
previously discussed in Ref.~\cite{NichettiZaccone2026Delayed};
the EPDM 4038 dataset is included as the linear-reference material
used in the present comparison. The experimental points show an
approximately common near-onset growth window followed by
material-dependent departure at larger deformation. The asymptotic
response-class annotations extend beyond the finite experimental
window and should not be interpreted as unique topology assignments
for every material.
\textbf{(b)} Viscous nonlinearity ratio (VLR) for EPDM 4038 and
NBR3330 at \SI{0.1}{Hz} and \SI{0.5}{Hz}. These data are used here
to compare the viscous nonlinear response of the linear-reference
and branched materials. VLR remains close to the experimental
resolution floor beyond the first nonlinear scale and begins a
clearly resolved increase only near
$\gamma_{\vix}>\gamma_0^{\LR}$. No normalized plateau or universal
VLR ceiling is imposed. The two panels emphasize that ELR and VLR
are post-processing observables of the same SPSD response but can
evolve differently with $\gamma_0$.}
\label{fig:experimental}
\end{figure*}

At each imposed SPSD amplitude, the signal is allowed to stabilize for approximately ten cycles. This harmonic-resolved strategy is consistent with the broader use of nonlinear rheology as an architecture-sensitive probe of branched polymer melts.~\cite{Auhl2008} Twenty-five subsequent cycles are acquired, with 1024 readings of imposed deformation and corresponding stress per cycle. The periodic stress is decomposed through the ninth harmonic. At \SI{0.1}{Hz}, each cycle lasts \SI{10}{s}, so the acquisition is correspondingly long.

The experimental estimator of $\gamma_0^{\LR}$ should be kept distinct from the theoretical scale in Eq.~\eqref{eq:onsetlin}. Operationally it is associated with the first experimentally resolvable higher odd harmonic above the instrumental and waveform noise floor. A formal statistical detection criterion is not defined here because the present measurements were not acquired with a dedicated cycle-to-cycle uncertainty protocol. ELR and VLR are then post-processing observables of the same total SPSD waveform. The later scale $\gamma_{\vix}$ marks macroscopically resolved VLR growth and does not define a second molecular transition.

Figure~\ref{fig:experimental} uses the supplied experimental figures directly rather than reconstructing the data. Panel (a) is the complete ELR compilation. Panel (b) retains only EPDM and NBR3330, which are also present in panel (a), so that the different evolution of ELR and VLR can be compared without introducing a new fitted or schematic viscous curve.

\section{Prospective transferability test in transient shear}
\label{sec:transient}

The time-domain character of GCCRM suggests, as a prospective extension beyond the present validation dataset, a test more stringent than SPSD alone: the same material-specific parameter set can be carried from periodic forcing into nonperiodic homogeneous shear without resetting the internal variables. This is a proposed validation program, not a completed result of the present paper. For a shear-rate change at time $t_s$, the molecular variables remain continuous,
\begin{equation}
\mathcal Z^{\mathrm{up}}(t_s^+)=\mathcal Z^{\mathrm{up}}(t_s^-),
\qquad
X_i(t_s^+)=X_i(t_s^-),
\label{eq:statecontinuity}
\end{equation}
while only the imposed forcing is changed. The state reached at the end of one deformation segment is therefore the initial condition for the next.

A first rotational-rheometer test should compare, with one parameter set per material, (i) startup from rest, (ii) step-up from an already established steady molecular state to a higher shear rate, and (iii) step-down while retaining the preceding state memory. For an ideal cone-and-plate geometry with radius $R$ and spatially uniform shear stress, the predicted torque follows directly from the total GCCRM stress,
\begin{equation}
T(t)=\frac{2\pi R^3}{3}\,\tau(t).
\label{eq:conetorque}
\end{equation}
Comparing a linear-reference EPDM with a branched NBR would test simultaneously the transfer from SPSD to transient shear and the architecture dependence of buildup and relaxation.

A second transferability test can use capillary step-shear protocols. In the OSCAR reference experiment, a well-characterized polystyrene melt at $180~{}^{\circ}\mathrm{C}$ was first held for five minutes at $20~\mathrm{s}^{-1}$, stepped to $200~\mathrm{s}^{-1}$, allowed to reach a new wall-stress plateau, and then stepped back to $20~\mathrm{s}^{-1}$; the time-dependent wall stress was obtained from the capillary pressure gradient.~\cite{Nichetti2019OSCAR} The corresponding measurement bridge may be written in the form
\begin{equation}
\tau_w(t)=\frac14\left[\frac{\dd P}{\dd(L/D)}\right]_t,
\label{eq:capillarybridge}
\end{equation}
with the same experimental pressure corrections required by the capillary protocol. A GCCRM capillary validation would couple the same local molecular state equations to the capillary kinematics and pressure measurement bridge; it would not introduce a capillary-specific constitutive law. The molecular state should remain continuous across flow-rate steps, and the resulting wall-stress/pressure transient should be compared with experiment. Such cross-protocol testing would directly address whether one upstream molecular dynamics can describe SPSD, startup, step-up, and step-down responses across geometries.

\section{Energetic consistency and scope}
\label{sec:thermo}

The configurational/dissipative terminology is used with a deliberately limited thermodynamic claim, consistent with the general nonequilibrium-thermodynamic requirement that constitutive evolution and dissipation be assessed together rather than inferred from nomenclature alone.~\cite{GrmelaOttinger1997,OttingerGrmela1997} Several substructures of GCCRM are compatible with standard non-negative-dissipation constructions, but a complete pointwise thermodynamic proof of the final mapped model is not claimed here. The explicit dissipative pathway has non-negative instantaneous power for the admissible parameter domain because Eq.~\eqref{eq:viscstress} gives $\tau_V\dot\gamma\ge0$. A complete pointwise proof of
\begin{equation}
\mathcal D=\bm\tau:\bm D-\dot\Psi\ge0
\label{eq:dissineq}
\end{equation}
for the merged GCCRM would additionally require an independently justified stored-energy representation of the constraint-limited mapping and a thermodynamic force governing $\nu_i$. The tanh--artanh mapping is a constitutive closure, not a microscopic free-energy derivation, and Eq.~\eqref{eq:nu} is a kinetic closure, not a gradient-flow law derived from a specified free energy. Appendix~\ref{app:thermo} records the consistency audit and possible mathematical completions without promoting them to established microscopic physics.

The same caution applies to scope. The current paper develops and checks GCCRM for homogeneous simple shear under SPSD in periodic steady state. It does not claim a complete arbitrary-three-dimensional constitutive theory.

\section{Discussion}

The central result of the GCCRM framework lies in its hierarchy rather than in any single reduced curve. Architecture and the common upstream molecular dynamics are constructed once. A direct configurational stress and a driven dissipative internal-variable stress are evaluated from that upstream state, recombined, and only then analyzed spectrally. This organization also removes a potential conceptual error: the configurational pathway is not physically identical to $G_n'$, and the dissipative pathway is not physically identical to $G_n''$. Both time-domain contributions can populate both Fourier quadratures; only the total SPSD waveform defines the experimental harmonic set.

The branch-point partition is likewise  placed at the natural logical level. Local three-leg force balance and local displacement compatibility do not by themselves generate $\Phi_{bb}^0$. The partition follows only after the explicit Gaussian coarse-grained compliance assumption and RCC homogenization over the $N_a$ long arms of the molecular architecture. The same construction shows that $\Phi_i^0$ carries the equilibrium configurational displacement/energy partition, whereas $\Phi_i(t)$ is used as the finite-deformation constraint attenuation in the nonlinear mapping; this is why the configurational stress does not contain a second factor $w_i$.

The linearization in Sec.~\ref{sec:linear} provides an important consistency check. The two-population configurational response generates both storage and loss and sums to $G_e$ in the high-frequency modal normalization, while $\eta_f$ represents only residual/fast modes outside those two conformation modes. This definition prevents the residual fast channel from double counting the same entangled modes at linear order. The normalized partition $\varphi_{\eta,i}$ specifies how that channel is distributed between populations, and the nonlinear activity $X_i$ then modulates it. CCR, stretch relaxation, and tube dilation appear in both the common molecular dynamics and the driver $\mathcal R_i$, but in different roles: first as state-relaxation mechanisms, second as activity measures feeding irreversible momentum transfer. This non-redundancy is a constitutive bookkeeping assumption that is made explicit and remains open to a deeper microscopic energetic derivation.

The reduced nonlinear relations are separated from the primary dynamics. Equations~\eqref{eq:Qlinsmooth}--\eqref{eq:Qbrsmooth} are smooth reduced closures used to summarize the ELR response classes and to generate the single-panel Fig.~\ref{fig:smoothregimes} with explicitly reported illustrative parameters. The crossover center $u_r$ is linked to the primary arm-retraction dynamics through Eq.~\eqref{eq:urhazard}; the remaining memory-shape parameters are still reduced closures. The approximate linearity with $\log_{10}(\gamma_0/\gamma_0^{\LR})$ belongs only to an intermediate window. The branched maximum is a derived consequence of the specified memory-survival closure and is therefore parameter dependent; the linear $\log_{10}(4)$ ceiling is a separate high-strain harmonic closure rather than a theorem following from the tanh orientation bound. On the viscous side, any logarithmic VLR law remains conditional on an intra-cycle waveform-shape closure, and bounded $X_i$ alone does not determine a nonzero VLR plateau.

A quantitative GCCRM validation must be performed from the primary time-domain equations rather than from the reduced response-class maps. For each material, molecular architecture and linear-response parameters are fixed first. The remaining constitutive closure parameters are then either constrained independently or identified once on a declared calibration subset. With that parameter set frozen, the primary state equations are integrated to periodic steady state at additional amplitudes and frequencies, the total stress is formed in the time domain, and its Fourier coefficients, ELR, and VLR are calculated without using $Q_{\mathrm{lin}}$ or $R_a$ in the forward solution. In the present terminology, only this last no-refit comparison is called a prediction. The reduced relations of Sec.~V are therefore interpretive/asymptotic response maps: agreement of those maps with an experimental response class does not replace harmonic-level validation of the primary constitutive equations.

\subsection{Relation to established molecular constitutive models}
GCCRM does not claim CCR, chain stretch, dynamic tube dilution, activated arm retraction, or Fourier analysis as new individual mechanisms. Rolie--Poly provides a reduced molecular constitutive description of reptation, stretch and CCR for entangled linear melts; Pom--Pom introduces explicit long-chain-branch architecture and branch-controlled backbone dynamics; hierarchical branched-tube theories describe activated arm retraction, dynamic dilution and architecture-dependent relaxation in substantially greater detail.~\cite{LikhtmanGraham2003,McLeishLarson1998,MilnerMcLeish1997,vanRuymbeke2006} GCCRM uses these established physical ideas as constraints and background, but organizes them differently. Its specific construction introduces an RCC compliance partition, a flow-active constraint state that dynamically modifies the tube and the RCC attenuation, and a downstream dissipative activation driven entirely by the same upstream molecular history. The configurational and dissipative time-domain stresses are then recombined before the already-established Fourier post-processing is applied. The novelty claimed here therefore lies in this combined constitutive architecture and in its resulting testable SPSD stress waveform, rather than in any one of its individual molecular ingredients.

A decisive comparison with these established models must ultimately be quantitative: GCCRM should reproduce harmonic-resolved SPSD data with one parameter set and should be tested against transient protocols without introducing protocol-specific constitutive parameters.

Because a complete parameter-constrained harmonic and transient validation has not yet been performed, the present contribution should be read as a molecular constitutive framework with explicit governing equations, closures, analytical checks, and experimentally testable consequences. Appendix~\ref{app:hierarchy} lists the admissible parameter domain and the intended identification route. Quantitative predictive status requires fixing architecture and linear-response parameters independently, constraining the remaining nonlinear closure parameters without reusing the target observable, and then testing the same parameter set against multiple SPSD amplitudes and frequencies.

\section{Conclusions}

The Generalized Constraint-Controlled Rheological Model (GCCRM) provides a single molecular constitutive framework for the response of entangled linear and long-arm-branched polymer melts under sinusoidal periodic shear deformation (SPSD). At each instant, the same molecular state---describing chain conformation and stretch, flow-active constraints, convective constraint release (CCR), dynamic tube dilation, and, for branched polymers, activated arm retraction---determines two contributions to the stress. One is a configurational contribution associated with the evolving molecular structure, and the other is a dissipative contribution with its own internal relaxation dynamics. These two stresses are added in the time domain before the total stress is analyzed by Fourier decomposition. This distinction is important because the configurational and dissipative pathways should not be identified directly with the storage and loss Fourier quadratures.

Within the stated reduced closures, GCCRM also identifies characteristic strain-amplitude regimes of the nonlinear response. The strain $\gamma_0^{\mathrm{LR}}$ denotes the predicted onset of nonlinear rheological response: experimentally, it corresponds to the first departure from the linear viscoelastic regime that can be resolved through the appearance of higher odd harmonics above the instrumental noise floor. Beyond this onset, the elastic nonlinearity can display an approximately logarithmic growth regime. For long-arm-branched polymers, delayed loss of orientational memory associated with arm retraction can subsequently produce an architecture-dependent maximum followed by a decrease at larger deformation, whereas the corresponding linear-polymer response approaches a finite high-strain limit. The viscous nonlinearity develops differently: the strain $\gamma_{\vix}>\gamma_0^{\mathrm{LR}}$ marks the later point at which growth of the viscous nonlinearity ratio becomes clearly resolved experimentally. It should therefore be regarded as an observable crossover rather than as a second molecular transition.

The linear-response limit provides an additional consistency check. The configurational moduli of the backbone and arm populations satisfy
\[
\sum_i G_i^0 = G_e ,
\]
where $G_i^0$ is the linear configurational modulus of population $i$ and $G_e$ is the entanglement, or plateau, modulus. The remaining fast dissipative modes are represented separately by the residual viscosity. Elastic and viscous nonlinearity ratios (ELR and VLR), as well as Lissajous representations, are consequently derived observables of the total SPSD stress waveform rather than independent constitutive ingredients.

A stringent next step is to test whether the same material-specific GCCRM parameter set can be transferred across different shear protocols without changing the molecular architecture or resetting the internal state variables. The model should first be tested simultaneously against SPSD measurements of ELR and VLR and against startup and step-up transients in rotational shear, and subsequently against the OSCAR step-up and step-down protocol. Demonstrating that one parameter set can describe these different deformation histories would provide a substantially stronger test of the constitutive framework. Until such tests are completed, the present work should be regarded as establishing the GCCRM constitutive architecture, its explicit closures, and their analytical consequences.

\section*{Data Availability}

The experimental data underlying the figures in this study are available from the corresponding authors upon reasonable request. Data previously reported in the literature are identified and cited in the corresponding figure captions and references.

\section*{Acknowledgments}

A.Z. gratefully acknowledges funding from the European Union through Horizon Europe ERC Grant number 101043968 ``Multimech''.

\appendix

\section{Branch-point partition: local mechanics, compatibility, and RCC homogenization}
\label{app:partition}

The branch-point partition is a constitutive result rather than an identity. Two levels must be kept separate: the local mechanics of one physical T-junction and the homogenized mechanics of the representative constitutive cell (RCC).

Consider one physical branch point $B$ at which the incoming backbone leg $bb^{-}$, outgoing backbone leg $bb^{+}$, and one arm $a$ meet. Let the configurational tension vectors exerted on the node be $\bm t_{bb^-}$, $\bm t_{bb^+}$, and $\bm t_a$. Treating the branch point as an internal massless node on the constitutive time scale gives the local three-leg balance
\begin{equation}
\bm t_{bb^-}+\bm t_{bb^+}+\bm t_a=\bm 0.
\label{eq:localforce}
\end{equation}
This equation contains neither $N_a$ nor a backbone--arm partition coefficient and therefore cannot by itself generate Eq.~\eqref{eq:Phi0}.

Molecular connectivity provides an independent local compatibility condition. If $\bm u_B$ is the displacement of the common material node,
\begin{equation}
\bm u_{bb^-}(B)=\bm u_{bb^+}(B)=\bm u_a(B)=\bm u_B.
\label{eq:localcompat}
\end{equation}
Equation~\eqref{eq:localcompat} does not require the finite backbone domain and the finite arm domain to possess the same coarse-grained end-to-end configurational displacement. Force balance and local compatibility therefore establish mechanical coupling but do not yet determine the RCC partition.

The next step is an explicit coarse-graining assumption. Let $q$ be the generalized configurational coordinate associated with the stress-bearing mode. For a Gaussian domain containing $n$ statistical segments of length $b$, the small-extension entropic free energy is
\begin{equation}
\mathcal F_G(q)=\frac{3k_BT}{2nb^2}q^2,
\label{eq:GaussianF}
\end{equation}
so the conjugate generalized force and compliance are
\begin{equation}
f=\frac{\partial\mathcal F_G}{\partial q}=\frac{3k_BT}{nb^2}q,
\qquad
q=C(n)f,
\qquad
C(n)=\frac{nb^2}{3k_BT}.
\label{eq:GaussianCompliance}
\end{equation}
Thus Gaussian configurational compliance scales linearly with the number of statistical units. GCCRM transfers this scaling, as an explicit coarse-grained closure, to stress-bearing entanglement strands,
\begin{equation}
C_i=C_eZ_i,
\label{eq:entCompliance}
\end{equation}
where $C_e$ is the effective compliance per entanglement strand for the chosen generalized coordinate. Equation~\eqref{eq:entCompliance} is a model assumption, not a general topological identity.

For the molecular architecture of Eq.~\eqref{eq:Mtotal}, the RCC contains $N_a$ equivalent long arms. Hence
\begin{equation}
C_a^{\mathrm{hom}}=N_aC_eZ_a,
\qquad
C_{bb}=C_eZ_{bb}.
\label{eq:compliancesRCC}
\end{equation}
Here $N_a$ is the number of long arms of the modeled molecule. It does not arise from the local three-leg force balance of a single T-junction; it enters only when the arm population is homogenized at RCC level.

The constitutive closure adopted at RCC level is a compliance-weighted redistribution of the stored generalized configurational displacement. Let $q_{bb}$ and $q_a$ be the portions stored in the backbone domain and homogenized arm population, respectively. With generalized multiplier $f^*$,
\begin{equation}
q_{bb}=C_{bb}f^*,
\qquad
q_a=C_a^{\mathrm{hom}}f^*,
\label{eq:qpartition}
\end{equation}
and cell-level compatibility gives
\begin{equation}
q=q_{bb}+q_a
=f^*\left(C_{bb}+C_a^{\mathrm{hom}}\right).
\label{eq:qtotal}
\end{equation}
The multiplier $f^*$ is a generalized coarse-grained conjugate force and is not identified with any one of the three local tension vectors in Eq.~\eqref{eq:localforce}. Defining $\Phi_{bb}^0=q_{bb}/q$ yields
\begin{equation}
\Phi_{bb}^{0}
=\frac{C_{bb}}{C_{bb}+C_a^{\mathrm{hom}}}
=\frac{Z_{bb}}{Z_{bb}+N_aZ_a}
=\frac{1}{1+\Xi},
\label{eq:PhiDerive}
\end{equation}
and normalization of the complementary stored fraction gives
\begin{equation}
\Phi_a^0=1-\Phi_{bb}^0
=\frac{N_aZ_a}{Z_{bb}+N_aZ_a}.
\label{eq:PhiADerive}
\end{equation}
The same compliance construction also fixes the stored generalized configurational-energy fractions. Since
\begin{equation}
\Psi_i=\frac12 f^*q_i=\frac12 C_i(f^*)^2,
\end{equation}
one obtains
\begin{equation}
\frac{\Psi_i}{\Psi_{bb}+\Psi_a}
=\frac{C_i}{C_{bb}+C_a^{\mathrm{hom}}}=\Phi_i^0.
\label{eq:energyPartition}
\end{equation}
Equation~\eqref{eq:energyPartition} establishes the configurational-displacement and quadratic-energy partition associated with the equilibrium RCC compliance closure. It does not imply that the dynamically updated $\Phi_i(t)$ remains an exact nonlinear stress fraction after application of the nonlinear constraint-limited orientation mapping. Applying $w_i$ again in the configurational stress would nevertheless double-weight the equilibrium backbone--arm architecture represented by the RCC closure. The mass fractions $w_i$ are retained where a physical population weighting is independently required.
The normalization $\Phi_{bb}^0+\Phi_a^0=1$ is an identity once complementary fractions have been defined; Eq.~\eqref{eq:PhiDerive} is not. It is conditional on local T-junction force balance, local displacement compatibility, the Gaussian coarse-grained compliance closure, and RCC homogenization over the $N_a$ arms.
\section{Onset, smooth elastic closures, maximum, and asymptotes}
\label{app:elasticregime}

The cooperative onset closure assumes that the resolved nonlinear source reaches a fixed molecular criterion when the coherent entanglement participation is multiplied by the squared strain amplitude,
\begin{equation}
Z_{bb}F_{\mathrm{CLF}}(Z_{bb})
\left(\gamma_{0,\mathrm{lin}}^{\LR}\right)^2=C_{\LR}^2.
\label{eq:onsetcriterion}
\end{equation}
This gives Eq.~\eqref{eq:onsetlin}. In the branched RCC, applying the same criterion to the equilibrium backbone participation gives
\begin{equation}
\Phi_{bb}^{0}
\left(\gamma_{0,\mathrm{br}}^{\LR}\right)^2
=\left(\gamma_{0,\mathrm{lin}}^{\LR}\right)^2,
\end{equation}
so that Eq.~\eqref{eq:onsetbr} follows. The square-root shift is therefore conditional on the cooperative-onset and branch-partition closures.

The smooth linear response class used in Fig.~\ref{fig:smoothregimes} follows Eq.~\eqref{eq:Qlinsmooth}. Its derivative is continuous for all $x>0$ and
\begin{equation}
Q_{\mathrm{lin}}(x)\rightarrow Q_{\infty}=3
\quad (x\rightarrow\infty),
\end{equation}
which yields $\mathrm{ELR}_{\mathrm{lin}}\rightarrow\log_{10}(4)$. The asymptote is therefore imposed by the high-strain harmonic closure $Q_{\infty}=3$; bounded orientation alone does not determine the numerical value.

For the branched response, Eq.~\eqref{eq:Rasmooth} is smooth for all $u$ and the maximum of Eq.~\eqref{eq:Qbrsmooth} is obtained from
\begin{equation}
\frac{\dd Q_{\mathrm{br}}}{\dd x}=0
\quad\Longleftrightarrow\quad
\frac{\dd\ln Q_{\mathrm{lin}}}{\dd\ln x}
+\frac{\dd\ln R_a}{\dd\ln x}=0,
\end{equation}
which is Eq.~\eqref{eq:peakcond}. Thus the maximum appears only after the quantitative memory closure has been specified; its position and height are not parameter-free. In a quantitative application its crossover center is linked to the primary activated arm-retraction time through Eq.~\eqref{eq:urhazard}.

For $u\gg u_r$,
\begin{equation}
R_a(u)\sim
\Xi^{-r_a\ln10/\kappa_a}
\exp[-r_a\ln10\,(u-u_r)].
\label{eq:RaAsymU}
\end{equation}
Since $x=10^u$,
\begin{equation}
R_a(x)\sim
\Xi^{-r_a\ln10/\kappa_a}
10^{r_au_r}x^{-r_a}.
\label{eq:RaAsymX}
\end{equation}
The branched reduced distortion therefore tends to zero as the linear reference saturates. This is the asymptote of the current smooth branch-memory closure, not a remnant of an earlier piecewise parameterization and not a geometrical construction of the figure.
\section{The dissipative activation as a causal fading-memory variable}
\label{app:Xmemory}

For $c_{\CCR},c_{\mathrm{st}},c_{\mathrm{td}}\ge0$, positive relaxation times, and the non-negative drivers in Eqs.~\eqref{eq:Rccr}--\eqref{eq:Rtd}, one has $\mathcal R_i(t)\ge0$. Write Eq.~\eqref{eq:Xdyn} as
\begin{equation}
\dot X_i+[a_i(t)+b_i]X_i=a_i(t),
\qquad
a_i(t)=k_{X,i}\mathcal R_i(t)\ge0,
\qquad
b_i=\frac{1}{\tau_{X,i}^{\rec}}>0.
\label{eq:Xlinear}
\end{equation}
For an initial time $t_0$, the exact solution is
\begin{equation}
X_i(t)=X_i(t_0)e^{-\int_{t_0}^{t}[a_i(s)+b_i]\dd s}
+\int_{t_0}^{t}a_i(u)
\exp\!\left[-\int_u^t[a_i(s)+b_i]\dd s\right]\dd u.
\label{eq:Xexact}
\end{equation}
After transients, the initial-condition term decays and
\begin{equation}
X_i(t)=\int_{-\infty}^{t}a_i(u)
\exp\!\left[-\int_u^t[a_i(s)+b_i]\dd s\right]\dd u.
\label{eq:Xfading}
\end{equation}
Thus $X_i$ has its own first-order kinetics but no independent molecular driver: it is a causal fading-memory functional of the past common-state forcing $\mathcal R_i$.

The interval $[0,1]$ is invariant. At $X_i=0$, $\dot X_i=a_i\ge0$; at $X_i=1$, $\dot X_i=-b_i<0$. Hence physical initial data remain physical.

If a prescribed driver $a_i(t)$ is periodic with period $T$, define
\begin{equation}
H_i(t)=\int_0^t[a_i(s)+b_i]\dd s.
\end{equation}
The periodic condition $X_i(T)=X_i(0)$ gives
\begin{equation}
X_i(0)=
\frac{\int_0^T a_i(u)e^{H_i(u)}\dd u}
{e^{H_i(T)}-1}.
\label{eq:Xperiodic}
\end{equation}
Because $b_i>0$, the denominator is positive. The difference between two solutions with the same prescribed driver decays as $e^{-H_i(t)}$. This proves uniqueness and stability of the periodic $X_i(t)$ \emph{conditional on the periodic driver $\mathcal R_i(t)$}. It does not prove uniqueness of the full coupled periodic state $\{\bm A_i,\nu_i,X_i\}$.
\section{Minimal bridge from dissipative activation to Fourier observables}
\label{app:fourierbridge}

The model-specific bridge is
\begin{equation}
X_i(t)\longrightarrow\eta_i(t)
\longrightarrow\tau_V(t)
\longrightarrow\text{Fourier projection},
\label{eq:bridgechain}
\end{equation}
not a direct map from an amplitude-domain $X_i(\gamma_0)$ to VLR. With $\theta=\omega t$ and $h_i(\theta)=X_i^q(\theta)$,
\begin{equation}
\frac{\tau_V(\theta)}{\gamma_0}
=\omega\eta_f\left[1+A_V\sum_i\varphi_{\eta,i}h_i(\theta)\right]\cos\theta.
\label{eq:tauVtheta}
\end{equation}
The diagnostic Fourier projections of this \emph{contribution} are
\begin{align}
G_{n,V}''&=\frac{\omega\eta_f}{\pi}
\int_0^{2\pi}\left[1+A_V\sum_i\varphi_{\eta,i}h_i(\theta)\right]\cos\theta\cos(n\theta)\,\dd\theta,
\label{eq:GppV}\\
G_{n,V}'&=\frac{\omega\eta_f}{\pi}
\int_0^{2\pi}\left[1+A_V\sum_i\varphi_{\eta,i}h_i(\theta)\right]\cos\theta\sin(n\theta)\,\dd\theta.
\label{eq:GpV}
\end{align}
Equations~\eqref{eq:GppV} and \eqref{eq:GpV} show explicitly that a phase-dependent viscosity can populate both quadratures. They are diagnostic projections of $\tau_V$ only; the experimental $G_n'$ and $G_n''$ are obtained from the total stress Eq.~\eqref{eq:totalstress}.

A cycle average such as $\overline X_{i,q}$ does not determine the higher harmonics because it discards the phase dependence of $h_i(\theta)$. A logarithmic amplitude-domain VLR relation can be obtained only with an additional waveform-shape closure. If, over a restricted asymptotic window,
\begin{equation}
h_i(\theta;\gamma_0,\omega)
\simeq A_X(\gamma_0,\omega)f_i(\theta;\omega),
\label{eq:shapeseparable}
\end{equation}
with a shape $f_i$ that varies weakly with $\gamma_0$, and if $A_X\propto(\gamma_0/\gamma_{\vix})^{rq}$, then the nonlinear Fourier amplitudes inherit the same amplitude factor. Only under this additional closure can one obtain conditionally
\begin{equation}
\mathrm{VLR}\simeq B_V+rq\log_{10}\left(\frac{\gamma_0}{\gamma_{\vix}}\right)
\label{eq:VLRconditional}
\end{equation}
over the corresponding window. This is an asymptotic reduced consequence, not a direct result of $0\le X_i\le1$.

The counterexample is immediate: if $X_i(t)\to1$ uniformly throughout the SPSD cycle, then the bracketed dissipative amplification tends to a constant. The dissipative contribution then contains no higher harmonics generated by viscosity modulation, so its nonlinear harmonic ratio tends back toward zero. A finite nonzero VLR ceiling would therefore require a nontrivial limiting intra-cycle waveform and cannot be specified without that additional information.
\section{Energetic consistency audit and possible completion}
\label{app:thermo}

For an isothermal incompressible melt the local requirement is
\begin{equation}
\mathcal D=\bm\tau:\bm D-\dot\Psi\ge0.
\label{eq:thermocriterion}
\end{equation}
Before application of the constraint-limited orientation mapping, a standard positive-definite conformation tensor admits the reference quadratic--logarithmic free-energy candidate
\begin{equation}
\Psi_i^{A,\mathrm{ref}}=\frac{G_e\Phi_i^0\nu_i}{2}
\left[\tr\bm A_i-\ln\det\bm A_i-3\right].
\label{eq:PsiA}
\end{equation}
This expression is used only as a reference energetic audit of the unconstrained conformation sector. It is not claimed to generate the final GCCRM configurational stress, because the latter contains the nonlinear, dynamically updated constraint-limited mapping $S_{xy,i}^{\CL}[\Phi_i(t),S_{xy,i}^{u}(t)]$. Consequently, Eq.~\eqref{eq:PsiA} cannot by itself establish pointwise thermodynamic consistency of the merged GCCRM. The viscous branch is explicitly non-negative,
\begin{equation}
\mathcal D_V=\eta_f\left[1+A_V\sum_i\varphi_{\eta,i}X_i^q\right]\dot\gamma^2\ge0.
\label{eq:DV}
\end{equation}

Two points are not closed by the present primary equations. First, Eq.~\eqref{eq:CLmap} is imposed directly as a functional orientation mapping. Because it is monotone, one can mathematically construct a potential whose derivative reproduces the mapped stress in simple shear. For example, with $q_i=\operatorname{artanh}(2S_{xy,i}^u)$, the convex function
\begin{equation}
f(s)=\frac14\left[(1+2s)\ln(1+2s)+(1-2s)\ln(1-2s)\right]
\end{equation}
has $f'(s)=\operatorname{artanh}(2s)$, and minimization of $f(s)-\Phi_iq_is$ reproduces the tanh--artanh relation. This establishes a possible mathematical energetic representation of the closure. It does \emph{not}, by itself, prove that the resulting potential is the physical orientational free energy stored by the entanglement network. Such an identification requires independent microscopic justification and is outside the present scope.

Second, even though Eq.~\eqref{eq:nu} has the correct quiescent fixed point, a full thermodynamic treatment would regard $\nu_i$ as an internal variable driven by a conjugate thermodynamic force. A possible completion is
\begin{equation}
\dot\nu_i=-M_{\nu,i}\frac{\partial\Psi}{\partial\nu_i},
\qquad M_{\nu,i}\ge0,
\label{eq:nugradient}
\end{equation}
which gives
\begin{equation}
\mathcal D_{\nu,i}=M_{\nu,i}
\left(\frac{\partial\Psi}{\partial\nu_i}\right)^2\ge0.
\end{equation}
Equation~\eqref{eq:nugradient} is presented as a thermodynamically admissible completion, not as an additional constitutive equation already established by the molecular GCCRM.

Consequently, the rigorous conclusion is limited but useful: several substructures of GCCRM are compatible with standard non-negative-dissipation constructions, but a complete pointwise thermodynamic proof of the final mapped model is not claimed here. What is established explicitly is non-negative instantaneous power of the dissipative pathway, together with the stated state-space invariance checks. A complete thermodynamic embedding of the final mapped configurational stress and of the $\nu_i$ kinetics remains an open constitutive completion. The thermodynamic layer constrains the architecture; it is not a third response branch and is not used to generate ELR or VLR.

\section{Model hierarchy, admissibility, and parameter identification}
\label{app:hierarchy}

A central point of the merged GCCRM is that not all relations have the same logical status. The classification below uses the categories definition/identity, governing equation, constitutive closure, asymptotic/reduced closure, derived consequence, and experimental observable. This distinction prevents reduced regime formulas or post-processing observables from being mistaken for the primary molecular dynamics.

\FloatBarrier
\begin{table*}[t]
\caption{Logical hierarchy of the merged GCCRM. Both constitutive pathways generate time-domain stresses; Fourier quadratures are obtained only after recombination into the total stress.}
\label{tab:hierarchy}
\small
\begin{tabular}{llll}
\parbox[t]{0.15\textwidth}{\textbf{Level}} &
\parbox[t]{0.27\textwidth}{\textbf{Principal relation}} &
\parbox[t]{0.27\textwidth}{\textbf{Status}} &
\parbox[t]{0.20\textwidth}{\textbf{Output}} \\[0.4em]
\hline\noalign{\vskip0.4em}
\parbox[t]{0.15\textwidth}{Architecture} &
\parbox[t]{0.27\textwidth}{$M_{bb}$, $M_a$, $N_a$, $Z_i$, $w_i$, $\Xi$} &
\parbox[t]{0.27\textwidth}{Definition/identity; $N_a$ is the long-arm count} &
\parbox[t]{0.20\textwidth}{Molecular loading variables} \\[0.7em]
\parbox[t]{0.15\textwidth}{Branch partition} &
\parbox[t]{0.27\textwidth}{$\Phi_{bb}^0$, $\Phi_a^0$} &
\parbox[t]{0.27\textwidth}{Constitutive closure after force balance, compatibility, compliance scaling, homogenization} &
\parbox[t]{0.20\textwidth}{Equilibrium constraint partition} \\[0.7em]
\parbox[t]{0.15\textwidth}{Upstream state} &
\parbox[t]{0.27\textwidth}{$\bm A_i$, $\lambda_i$, $\bm S_i^u$, $\nu_i$, $Z_i^{\eff}$, $\Phi_i$} &
\parbox[t]{0.27\textwidth}{Governing state plus explicit tube/orientation closures} &
\parbox[t]{0.20\textwidth}{Molecular trajectory generated by governing equations} \\[0.7em]
\parbox[t]{0.15\textwidth}{Upstream dynamics} &
\parbox[t]{0.27\textwidth}{Eqs.~\eqref{eq:CCR}, \eqref{eq:nu}, \eqref{eq:Acommon}, \eqref{eq:tauR}} &
\parbox[t]{0.27\textwidth}{Governing equations plus constitutive closures} &
\parbox[t]{0.20\textwidth}{Molecular state under the imposed shear history} \\[0.7em]
\parbox[t]{0.15\textwidth}{Config. pathway} &
\parbox[t]{0.27\textwidth}{Eq.~\eqref{eq:elasticstress}} &
\parbox[t]{0.27\textwidth}{Primary time-domain constitutive projection} &
\parbox[t]{0.20\textwidth}{$\tau_E(t)$} \\[0.7em]
\parbox[t]{0.15\textwidth}{Dissipative pathway} &
\parbox[t]{0.27\textwidth}{$\mathcal Z_i\to\mathcal R_i\to X_i\to\eta_i\to\tau_V$} &
\parbox[t]{0.27\textwidth}{Governing $X_i$ kinetics plus constitutive closures} &
\parbox[t]{0.20\textwidth}{$\tau_V(t)$} \\[0.7em]
\parbox[t]{0.15\textwidth}{Recombination} &
\parbox[t]{0.27\textwidth}{$\tau=\tau_E+\tau_V$} &
\parbox[t]{0.27\textwidth}{Constitutive decomposition defining total GCCRM shear stress} &
\parbox[t]{0.20\textwidth}{Total GCCRM shear stress} \\[0.7em]
\parbox[t]{0.15\textwidth}{Reduced regime map} &
\parbox[t]{0.27\textwidth}{Eqs.~\eqref{eq:Qlinsmooth}, \eqref{eq:Rasmooth}, \eqref{eq:Qbrsmooth}, \eqref{eq:Xpower}} &
\parbox[t]{0.27\textwidth}{Asymptotic/reduced closures} &
\parbox[t]{0.20\textwidth}{Response-class scalings and asymptotes} \\[0.7em]
\parbox[t]{0.15\textwidth}{Fourier bridge} &
\parbox[t]{0.27\textwidth}{Eq.~\eqref{eq:FourierStress} and published post-processing} &
\parbox[t]{0.27\textwidth}{Experimental/post-processing definition} &
\parbox[t]{0.20\textwidth}{$G_n'$, $G_n''$, ELR, VLR, Lissajous representation} \\[0.7em]
\parbox[t]{0.15\textwidth}{Thermo. audit} &
\parbox[t]{0.27\textwidth}{$\mathcal D=\bm\tau:\bm D-\dot\Psi\ge0$} &
\parbox[t]{0.27\textwidth}{Cross-cutting consistency requirement} &
\parbox[t]{0.20\textwidth}{Energetic admissibility conditions} \\[0.4em]
\end{tabular}
\normalsize
\end{table*}

\subsection{Parameter admissibility and physical state space}

The mathematical checks used in the manuscript require the admissible parameter domain
\begin{equation}
\begin{split}
&G_e,M_e,M_{bb},M_a,\eta_f,\tau_e,C_a>0,\\
&\tau_{\rel,i}^{0},\tau_{R,i}^{0},\tau_{\rec,i},\tau_{X,i}^{\rec}>0.
\end{split}
\end{equation}
\begin{align}
N_a&\in\mathbb N_0,\qquad q>0,\qquad \lambda_{\max,i}>1,\\
\alpha_i,\beta_i,\beta_{s,i},k_{X,i},A_V&\ge0,\qquad \mu_a,\mu_d\ge0,\qquad \eta_a>0,\\
c_{\CCR},c_{\mathrm{st}},c_{\mathrm{td}}&\ge0,\qquad c_{\CCR}+c_{\mathrm{st}}+c_{\mathrm{td}}=1,\\
\varphi_{\eta,i}&\ge0,\qquad \sum_i\varphi_{\eta,i}=1,\\
C_{\LR},F_{\mathrm{CLF}}(Z),x_s,m&>0,\qquad \kappa_a,r_a>0.
\label{eq:paramdomain}
\end{align}
The physical state space further requires $0\le\nu_i\le1$, $0\le X_i\le1$, $\bm A_i$ positive definite, and $0<\lambda_i<\lambda_{\max,i}$ for the SPSD states considered here. Invariance of the first two intervals is proved directly in the main text and Appendix~\ref{app:Xmemory}. Positive definiteness does not require $F_i(\lambda_i)\ge0$. The non-affine evolution can be written locally as a strictly positive isotropic source plus a scalar multiple of $\bm A_i$; at the boundary of the positive-definite cone the isotropic source remains positive even when the scalar coefficient changes sign during transient contraction. The affine part remains a congruence transformation \cite{Enzo}. The divergence of $F_i$ as $\lambda_i\to\lambda_{\max,i}^{-}$ provides a restoring finite-extensibility barrier. The present arm-retraction closure is restricted to entangled long arms: the branched implementation considered here assumes an entangled backbone, $Z_{bb}\gg1$, and $Z_a>1$ whenever the activated arm-retraction law is invoked. The limit of unentangled short side groups is outside the scope of this specific closure and would require a different local relaxation description. A global proof for arbitrary three-dimensional flows is not claimed; the present admissibility statement applies to the homogeneous simple-shear SPSD scope of the paper and should be enforced with structure-preserving numerical integration.

\subsection{Parameter identification hierarchy}

Table~\ref{tab:parameters} states how parameters should be constrained before a quantitative claim of prediction is made. Numerical values are material dependent and are not assigned in the present framework draft; this is deliberate because the required single-parameter-set harmonic validation has not yet been completed.

\begin{table*}[t]
\caption{Parameter-identification hierarchy. ``Not assigned here'' means that the present manuscript specifies the role and admissible source of a parameter but does not report a fitted numerical value.}
\label{tab:parameters}
\small
\begin{tabular}{llll}
\parbox[t]{0.18\textwidth}{\textbf{Parameter(s)}} &
\parbox[t]{0.15\textwidth}{\textbf{Dimension}} &
\parbox[t]{0.27\textwidth}{\textbf{Physical role}} &
\parbox[t]{0.30\textwidth}{\textbf{Identification route/status}} \\[0.4em]
\hline\noalign{\vskip0.4em}
\parbox[t]{0.18\textwidth}{$M_e,M_{bb},M_a,N_a$} &
\parbox[t]{0.15\textwidth}{mass; dimensionless counts} &
\parbox[t]{0.27\textwidth}{Molecular architecture and long-arm population} &
\parbox[t]{0.30\textwidth}{Molecular characterization/architecture model where available; $N_a$ is the long-arm count. Not assigned here.} \\[0.7em]
\parbox[t]{0.18\textwidth}{$G_e,\tau_{\rel,i}^{0}$} &
\parbox[t]{0.15\textwidth}{Pa; s} &
\parbox[t]{0.27\textwidth}{Two-population reduced linear configurational response} &
\parbox[t]{0.30\textwidth}{Must be constrained from linear rheology before nonlinear SPSD fitting.} \\[0.7em]
\parbox[t]{0.18\textwidth}{$\eta_f$, $\varphi_{\eta,i}$} &
\parbox[t]{0.15\textwidth}{Pa s; dimensionless} &
\parbox[t]{0.27\textwidth}{Residual/fast-mode linear dissipation and its normalized population partition} &
\parbox[t]{0.30\textwidth}{$\eta_f$ is constrained from residual $G''(\omega)$ after the two-population conformation response is fixed, within the window $\omega\tau_{\mathrm{fast,max}}\ll1$; $\varphi_{\eta,i}$ is fixed by the minimal closure $\varphi_{\eta,i}=w_i$ here or by independent information in a future extension.} \\[0.7em]
\parbox[t]{0.18\textwidth}{$\alpha_i,\mu_d$} &
\parbox[t]{0.15\textwidth}{dimensionless} &
\parbox[t]{0.27\textwidth}{Dynamic tube-dilation scaling} &
\parbox[t]{0.30\textwidth}{Constitutive/scaling closure; constrain independently where possible. Not assigned here.} \\[0.7em]
\parbox[t]{0.18\textwidth}{$\tau_e,C_a,\mu_a,\eta_a$} &
\parbox[t]{0.15\textwidth}{s; dimensionless} &
\parbox[t]{0.27\textwidth}{Activated arm-retraction scaling} &
\parbox[t]{0.30\textwidth}{Molecular/linear-relaxation information and prior arm-retraction scaling. Not assigned here.} \\[0.7em]
\parbox[t]{0.18\textwidth}{$\beta_i,\tau_{\rec,i}$} &
\parbox[t]{0.15\textwidth}{dimensionless; s} &
\parbox[t]{0.27\textwidth}{CCR efficiency and constraint recovery} &
\parbox[t]{0.30\textwidth}{Nonlinear common-state closure; constrain without reusing target ELR/VLR where possible.} \\[0.7em]
\parbox[t]{0.18\textwidth}{$\beta_{s,i},\lambda_{\max,i}$} &
\parbox[t]{0.15\textwidth}{dimensionless} &
\parbox[t]{0.27\textwidth}{Flow-assisted stretch relaxation and finite extensibility} &
\parbox[t]{0.30\textwidth}{Constitutive closure/molecular extensibility estimate. Not assigned here.} \\[0.7em]
\parbox[t]{0.18\textwidth}{$c_{\CCR},c_{\mathrm{st}},c_{\mathrm{td}}$, $k_{X,i},\tau_{X,i}^{\rec}$, $A_V,q$} &
\parbox[t]{0.15\textwidth}{mixed dimensionless/s} &
\parbox[t]{0.27\textwidth}{Normalized dissipative-channel mixture, activation rate, and nonlinear viscous amplification} &
\parbox[t]{0.30\textwidth}{Impose $c_{\CCR}+c_{\mathrm{st}}+c_{\mathrm{td}}=1$ so $k_{X,i}$ is the overall activation-rate scale; remaining nonlinear parameters require harmonic-level identification/validation. Not assigned here.} \\[0.7em]
\parbox[t]{0.18\textwidth}{$C_{\LR},F_{\mathrm{CLF}},x_s,m$, $u_r,\kappa_a,r_a$} &
\parbox[t]{0.15\textwidth}{dimensionless} &
\parbox[t]{0.27\textwidth}{Reduced onset and ELR regime-map closures} &
\parbox[t]{0.30\textwidth}{Reduced/asymptotic description; not primary time-domain parameters and not evidence of prediction unless fixed independently.} \\[0.4em]
\end{tabular}
\normalsize
\end{table*}

\subsection{Illustrative parameters used to render Fig.~\ref{fig:smoothregimes}}

Figure~\ref{fig:smoothregimes} is a single-panel schematic response-class map. Its numerical values are reported only to make the graphic reproducible; they are not material fits and are not used as evidence of quantitative prediction. The linear curve is generated by Eq.~\eqref{eq:Qlinsmooth}, and the branched curves by Eqs.~\eqref{eq:Rasmooth} and \eqref{eq:Qbrsmooth}.

\begin{table*}[t]
\caption{Illustrative numerical parameters used to render Fig.~\ref{fig:smoothregimes}. In a quantitative material calculation $u_r$ should be obtained from the cycle-integrated arm-retraction matching condition in Eq.~\eqref{eq:urhazard}; the values shown here are graphical response-class choices only.}
\label{tab:fig2params}
\small
\begin{tabular}{llll}
\parbox[t]{0.17\textwidth}{\textbf{Class}} &
\parbox[t]{0.12\textwidth}{\textbf{Parameter}} &
\parbox[t]{0.09\textwidth}{\textbf{Value}} &
\parbox[t]{0.49\textwidth}{\textbf{Role}} \\[0.3em]
\hline\noalign{\vskip0.35em}
\parbox[t]{0.17\textwidth}{all} & $Q_{\infty}$ & $3.000$ & \parbox[t]{0.50\textwidth}{High-strain harmonic closure, giving $\log_{10}(4)$.} \\[0.4em]
\parbox[t]{0.17\textwidth}{all} & $x_s$ & $3.80$ & \parbox[t]{0.50\textwidth}{Midpoint scale of smooth backbone buildup.} \\[0.4em]
\parbox[t]{0.17\textwidth}{all} & $m$ & $1.60$ & \parbox[t]{0.50\textwidth}{Smooth backbone crossover exponent.} \\[0.4em]
\parbox[t]{0.17\textwidth}{sparse branched} & $\Xi$ & $0.40$ & \parbox[t]{0.50\textwidth}{Illustrative arm loading.} \\[0.4em]
\parbox[t]{0.17\textwidth}{sparse branched} & $u_r$ & $0.504$ & \parbox[t]{0.50\textwidth}{Illustrative memory-loss center; quantitative value constrained by Eq.~\eqref{eq:urhazard}.} \\[0.4em]
\parbox[t]{0.17\textwidth}{sparse branched} & $\kappa_a$ & $1.333$ & \parbox[t]{0.50\textwidth}{Memory-crossover sharpness.} \\[0.4em]
\parbox[t]{0.17\textwidth}{sparse branched} & $r_a$ & $0.196$ & \parbox[t]{0.50\textwidth}{High-strain memory-loss exponent.} \\[0.4em]
\parbox[t]{0.17\textwidth}{stronger branched} & $\Xi$ & $2.00$ & \parbox[t]{0.50\textwidth}{Illustrative arm loading.} \\[0.4em]
\parbox[t]{0.17\textwidth}{stronger branched} & $u_r$ & $0.441$ & \parbox[t]{0.50\textwidth}{Illustrative memory-loss center; quantitative value constrained by Eq.~\eqref{eq:urhazard}.} \\[0.4em]
\parbox[t]{0.17\textwidth}{stronger branched} & $\kappa_a$ & $20.0$ & \parbox[t]{0.50\textwidth}{Memory-crossover sharpness.} \\[0.4em]
\parbox[t]{0.17\textwidth}{stronger branched} & $r_a$ & $0.306$ & \parbox[t]{0.50\textwidth}{High-strain memory-loss exponent.} \\[0.2em]
\end{tabular}
\normalsize
\end{table*}

The thermodynamic layer is cross-cutting rather than a third constitutive pathway. Non-negative dissipation constrains both projections and the common state, but the current tanh--artanh mapping and constraint-recovery kinetics remain constitutive closures rather than microscopically derived thermodynamic laws.


\clearpage
\bibliography{gccrm_jor_refs}

\end{document}